\documentclass[prd,singlecolumn,amsmath,nobibnotes,nofootinbib,superscriptaddress,preprintnumbers,floatfix]{revtex4}

\usepackage{amsmath,amssymb}
\usepackage{graphicx}
\usepackage[usenames, dvipsnames]{color}
\usepackage{units}
\usepackage{bm}

\newcommand{\ba}{\begin{eqnarray}}
\newcommand{\ea}{\end{eqnarray}}

\newcommand{\h}{\tilde{h}}

\begin{document}

\begin{flushleft}
KCL-PH-TH/2017-44
\end{flushleft}
\vskip.5truecm
\title{Neutron star mergers as a probe of modifications of general relativity with finite-range scalar forces}

\author{Laura Sagunski}
  \email{sagunski@yorku.ca}
  \affiliation{Department of Physics and Astronomy, York University, Toronto, Ontario, M3J 1P3, Canada}
  \affiliation{Perimeter Institute for Theoretical Physics, Waterloo, Ontario N2L 2Y5, Canada}
  
  \author{Jun Zhang}
  \email{jun34@yorku.ca}
  \affiliation{Department of Physics and Astronomy, York University, Toronto, Ontario, M3J 1P3, Canada}
  \affiliation{Perimeter Institute for Theoretical Physics, Waterloo, Ontario N2L 2Y5, Canada}

\author{Matthew C. Johnson}
  \email{mjohnson@perimeterinstitute.ca}
  \affiliation{Department of Physics and Astronomy, York University, Toronto, Ontario, M3J 1P3, Canada}
  \affiliation{Perimeter Institute for Theoretical Physics, Waterloo, Ontario N2L 2Y5, Canada}

\author{Luis Lehner}
  \email{llehner@perimeterinstitute.ca}
  \affiliation{Perimeter Institute for Theoretical Physics, Waterloo, Ontario N2L 2Y5, Canada}

  \author{Mairi Sakellariadou}
  \email{mairi.sakellariadou@kcl.ac.uk}
   \affiliation{Perimeter Institute for Theoretical Physics, Waterloo, Ontario N2L 2Y5, Canada}
 \affiliation{Theoretical Particle Physics and Cosmology Group, Physics Department, King's College London, University of London, Strand, London WC2R 2LS, UK}

  \author{Steven L. Liebling}
  \affiliation{Long Island University, Brookville, New York 11548, USA}

  \author{Carlos Palenzuela}
  \affiliation{Departament de Fisica, Universitat de les Illes Balears and Institut d'Estudis Espacials e Catalunya, Palma de Mallorca, Baleares E-07122, Spain}

  \author{David Neilsen}
  \affiliation{Department of Physics and Astronomy, Brigham Young University, Provo, Utah 84602, USA}

\date{\today}

\begin{abstract}
Observations of gravitational radiation from compact binary systems provide an unprecedented opportunity to test General Relativity in the strong field dynamical regime. In this paper, we investigate how future observations of gravitational radiation from binary neutron star mergers might provide constraints on finite-range forces from a universally coupled massive scalar field. Such scalar degrees of freedom are a characteristic feature of many extensions of General Relativity. For concreteness, we work in the context of metric $f(R)$ gravity, which is equivalent to General Relativity and a universally coupled scalar field with a non-linear potential whose form is fixed by the choice of $f(R)$. In theories where neutron stars (or other compact objects) obtain a significant scalar charge, the resulting attractive finite-range scalar force has implications for both the inspiral and merger phases of binary systems. We first present an analysis of the inspiral dynamics in Newtonian limit, and forecast the constraints on the mass of the scalar and charge of the compact objects for the Advanced LIGO gravitational wave observatory. We then perform a comparative study of binary neutron star mergers in General Relativity with those of a one-parameter model of $f(R)$ gravity using fully relativistic hydrodynamical simulations. These simulations elucidate the effects of the scalar on the merger and post-merger dynamics. We comment on the utility of the full waveform (inspiral, merger, post-merger) to probe different regions of parameter space for both
the particular model of $f(R)$ gravity studied here and for finite-range scalar forces more generally. 
\end{abstract}

\maketitle


\section{Introduction}

The recent detection of gravitational waves by the Advanced LIGO detectors~\cite{TheLIGOScientific:2016pea,Abbott:2016nmj,Abbott:2016blz} has confirmed a cornerstone of Einstein's theory of General Relativity (GR). Modifications of GR have already been tightly constrained by solar-system tests~\cite{Will:2014kxa} and observations of binary pulsars~\cite{Stairs:2003eg}, but such tests are restricted to the regime in which low order post-Newtonian calculations accurately describe the dynamics.  The detection of gravitational waves produced, for instance, by binary compact object mergers Indeed, the Advanced LIGO observations of the coalescing binary black holes GW150914~\cite{Abbott:2016blz}, GW151226~\cite{Abbott:2016nmj}, LVT151012 and GW170104~\cite{Abbott:2017vtc} have already led to some important constraints on deviations from GR~\cite{TheLIGOScientific:2016src,Yunes:2016jcc}. Recently, a Bayesian method~\cite{Callister:2017ocg} has been presented by which Advanced LIGO would detect or constrain the presence of vector and scalar polarizations in the stochastic background, hence providing a test of GR in the strong-field highly dynamical regime. Furthermore, the addition of Advanced Virgo, even though it does not significantly improve our detection ability, it may considerably improve our ability to estimate the parameters of backgrounds of mixed polarization~\cite{Callister:2017ocg}.
The non-detection of extra polarizations in the stochastic background may lead to experimental constraints on modified gravity theories, some of which may allow for up to four additional polarization modes. 

A major obstacle in further constraining modified gravity theories through gravitational wave observations is the lack of  a comprehensive understanding of dynamics in the strong-field highly dynamical regime. To contribute to filling this gap, we undertake the study of compact binary mergers in $f(R)$ theories (see Ref.~\cite{DeFelice:2010aj} for a review), as they represent one of the simplest modifications to GR. 
Note that $R$ stands for the Ricci scalar and $f(R)$ denotes an arbitrary function of $R$. In $f(R)$ gravity, the initial value problem is
typically well-posed, and therefore amenable to studying the non-linear evolution of
relevant initial configurations. 
Many other modified gravity theories quite likely do not define well-posed initial value problems, in particular those with higher derivatives or those with a complicated non-linear derivative structure, as found in --for
instance--  massive gravity~\cite{Bergshoeff:2009zz,deRham:2010kj}, bigravity~\cite{Hassan:2011tf}, 
Chern-Simons gravity~\cite{Alexander:2009tp}, non-commutative geometry~\cite{Connes:1994yd} (to name just
a few) and where, beyond clear general statements that one can make at the mathematical level,  
specific studies have explicitly indicated major
reasons for concern~\cite{Brito:2014ifa,Delsate:2014hba,Papallo:2017qvl,Cayuso:2017iqc}\footnote{
Considering the spectral action of an almost commutative torsion geometry, it has been shown~\cite{Sakellariadou:2016dfl} that the obtained Hamiltonian is
bounded from below, hence non-commutative spectral geometry, a theory that offers a purely geometric explanation for the Standard Model of particle physics~\cite{Chamseddine:2006ep},  does not suffer from linear instability.}.
To date there is no definitive well defined and consistent way to straightforwardly access predictions for gravitational wave observables for
general modifications in the highly dynamical, strong-gravity regime. Such a question is beginning to receive attention and
solutions in certain contexts are being proposed, see~\cite{Cayuso:2017iqc} for a solution in certain contexts,
see also~\cite{Campanelli:1994sj,Okounkova:2017yby}. Therefore, one motivation for the study of $f(R)$ gravity is that the relevant physics can be simulated with the tools at hand, and the observed phenomenology can be used as guidance towards possible deviations to consider.

Another motivation is the role that such theories might play in cosmology~\cite{Faulkner:2006ub}. One of the first successful models of cosmic inflation was based on $f(R)$ gravity~\cite{1980PhLB...91...99S}, which can account for the early inflationary era in the absence of a scalar field with a suitably fine-tuned potential, which is  used in scalar-field models of inflation. More precisely, the model $f(R)=R+a_2 R^2$ (with $a_2>0$, hence it does not suffer from the appearance of ghosts) leads to an exponential expansion driven by the $a_2 R^2$ term. Subsequently, $f(R)$ gravity was employed for infrared modifications of gravity.
Instead of introducing a mysterious dark energy component, $f(R)$ gravity was proposed as a dynamical explanation of the present cosmic acceleration~\cite{Boisseau:2000pr,EspositoFarese:2000ij,Capozziello:2003tk,Carroll:2003wy,Hu:2007nk}.  Such phenomenological proposals are valid provided certain constraints are imposed on the function $f(R)$, for the model at hand to be linearly stable and cosmologically viable.

Under certain conditions, and certainly in the cases we focus here, it can be shown that  $f(R)$ gravity is dynamically equivalent to Einstein gravity minimally coupled to a scalar field with a non-trivial potential that couples to matter through the trace of the energy momentum tensor in the Einstein frame~\cite{Sotiriou:2006hs}. To the extent that the scalar field is light, it sources a (finite-range) fifth force that couples universally, endowing objects with a scalar charge. We can therefore consider $f(R)$ gravity as a proxy for other theories with a finite-range scalar force. Any theory with such a fifth force is constrained~\cite{Fischbach:1999bc} by laboratory searches~\cite{Adelberger:2009zz,Kapner:2006si}, solar system tests~\cite{DeFelice:2009aj}, and cosmology~\cite{DeFelice:2008wz}. In some cases, the fifth force can be screened, and such constraints evaded, for example through the chameleon mechanism~\cite{Khoury:2003rn}. This mechanism generates an effective mass for the scalar field, through the interplay of scalar field self-interactions and interactions with the ambient 
matter (see however~\cite{Erickcek:2013oma}).

In $f(R)$ gravity, there can be non-trivial implications for the structure of compact objects such as neutron stars~\cite{Yazadjiev:2016pcb,Ramazanoglu:2016kul,Mendes:2014ufa,Doneva:2015hsa,Pani:2014jra,Yunes:2009ch,Frolov:2008uf,Upadhye:2009kt,Cooney:2009rr}, implying a possible window into gravity and cosmology through the study of compact objects. The study of binary neutron star or neutron star-black hole systems in general scalar tensor theories has already yielded a number of important results. For example, some theories exhibit the phenomenon of spontaneous scalarization~\cite{0264-9381-9-9-015,PhysRevLett.70.2220}, in which the scalar charge of neutron stars is environmentally dependent, and therefore can evolve during the merger~\cite{Palenzuela:2013hsa,Shibata:2013pra,Barausse:2012da,Sampson:2014qqa}. Such observed behavior underlines the
importance of exploring the full dynamics of theories of interest to reveal 
unanticipated behavior. In addition, the orbital dynamics in neutron star-black hole~\cite{Yunes:2011aa,Cardoso:2011xi} and other binaries~\cite{Alsing:2011er} can be affected in interesting ways. Finally, scalar charged neutron stars can arise in other motivated scenarios, including for some region of parameter space for the QCD axion~\cite{Hook:2017psm}, with implications for binary neutron star mergers.

While black hole mergers are an excellent laboratory to test some extensions of GR, they are not useful for testing $f(R)$ gravity because black holes cannot carry a scalar charge, and therefore the merger history would be exactly as predicted in GR (see however~\cite{Horbatsch:2011ye,Canate:2015dda}). Thus, because of their relatively large compaction, neutron stars are the most viable distribution of matter that can be used to probe $f(R)$ gravity in the strong-field regime, and so in this paper we focus on binary neutron star mergers. 

During the inspiral phase, once the neutron stars are within the Compton wavelength of the scalar, the scalar force accelerates the merger. This leads to a potentially observable modification to the gravitational wave form during the inspiral phase. During the merger, the scalar could in principle affect the final state (e.g. whether for a given system the merger results in a final state neutron star or prompt collapse into a black hole) and the post-merger dynamics. Such post-merger dynamics can lead to a long-lived time-dependent quadrupole, giving rise to a detectable post-merger gravitational wave signature (see e.g.~\cite{Lehner:2014asa} for a review). In principle, more total energy can be emitted in gravitational radiation after the merger than during the inspiral~\cite{Hanna:2016uhs}, implying that the post-merger waveform can provide important constraints as well. Constraints from the post-merger phase, however, have significant uncertainties associated with the neutron star equation of state, the role of magnetic fields and other forms of dissipation, etc. Nevertheless, one may hope that the observation of electromagnetic counterparts can assist in minimizing these astrophysical uncertainties (see e.g.~\cite{Ponce:2014hha}). Finally, although we focus on a particular model of $f(R)$ gravity for concreteness, our results for the inspiral, and to some degree our results for the post-merger phase, are generally applicable to any theory giving rise to a universally coupled massive scalar field. This is because any theory of $f(R)$ gravity in the Einstein frame can be approximated by a universally coupled massive scalar field in the linear regime relevant to the inspiral, and as long as the non-linearities in the scalar sector are not probed in the merger.

In this paper, we take a combined analytic and numerical approach. Because it is too computationally expensive to simulate a significant portion of the inspiral, and because the dynamics are well-described analytically, in Sec.~\ref{Sec:Newton}, we present a model of the inspiral dynamics in the presence of a universally coupled massive scalar in the Newtonian approximation. We find the associated gravitational waveform and estimate the potential constraints on the range of the force and the scalar charges that can be obtained by Advanced LIGO at design sensitivity.  The Newtonian model cannot accurately capture the merger and post-merger phases of the evolution, which additionally can in principle depend on the structure of the full theory, including non-linearities in the gravitational and scalar sectors. We therefore specialize to numerical simulations in $f(R)$ gravity. We outline the theory in Sec.~\ref{Sec:metricfR}, and present the results from fully relativistic hydrodynamical simulations of neutron star binary mergers in Sec.~\ref{sec:numerical}, first presenting individual neutron star solutions and then mergers. We conclude in Sec.~\ref{Sec:conclusions}. We work in natural units with $c=\hbar=1$.

{\bf Note:} Shortly after the first version of this paper was submitted to the arXiv, the first detection of gravitational waves from a binary neutron star merger was announced by LIGO~\cite{PhysRevLett.119.161101}. This event makes the scenario presented in this paper imminently testable with existing data, an effort that we hope to undertake in follow-up work. 

\section{Inspiral in the presence of a finite-range scalar force}\label{Sec:Newton}

Before specializing to $f(R)$ gravity in the fully dynamical regime, we first study the inspiral dynamics of a binary system in the presence of gravity and a generic finite-range scalar force at Newtonian order. This will provide a qualitative understanding of the dynamics and the interesting regions of parameter space, but will miss the effects of non-linear dynamics that become increasingly important late in the inspiral. To explore this regime, we rely on the results of our simulations, presented in Sec.~\ref{sec:numerical}. Here, we begin by reviewing the mechanism by which neutron stars acquire a scalar charge, then calculate the orbital dynamics along with the associated gravitational waveform, and finally forecast constraints on the range of the scalar force and charge of neutron stars that could be possible in the near-future with Advanced LIGO.

\subsection{Scalar charged neutron stars}

In this paper, we consider scenarios where a massive scalar $\phi$ with potential $V$, couples linearly to the trace of the energy momentum tensor. Neglecting gravity, the action is generically 
\begin{equation}
S = - \int d^4x \left[ \frac{\partial \phi^2}{2}  + V - \beta \frac{\phi}{M_{\rm Pl}} T^{\mu}_{\mu}  \right] \, .
\end{equation}
In this section, we assume a spherically symmetric flat spacetime with radial coordinate $r$. As a starting point, we neglect all relativistic effects and approximate the matter energy-momentum tensor, which is sourced by the neutron star, as 
\begin{equation}
  T^{\mu}_{\mu} =-\rho + 3 p  \simeq -\rho \,.
  \label{Tappr}
\end{equation}
Here, $\rho$ and $p$ are respectively the matter density and pressure of the neutron star, and we assume that the density field is static. For a static scalar field, which we will denote by $\varphi_{0}(r)$, the equation of motion is
\begin{equation}
\frac{d^2 \varphi_{0}}{dr^2} + \frac{2}{r}\frac{d\varphi_{0}}{dr} = 
m^2 \varphi_{0}
+
\frac{\beta}{M_{\rm Pl}} \rho\,,
\label{EoMvarphi}
\end{equation}
where the dimensionless parameter $\beta$ characterizes the strength of the coupling between the scalar and energy momentum tensor, $m$ stands for the mass of the scalar field and $M_{\rm Pl}^2 \equiv (8 \pi G)^{-1}$. In the $f(R)$ theories studied below, $\beta = 6^{-1/2}$; this can be kept in mind as a representative value in the following. We can solve this inhomogeneous differential equation by using a Green's function
\begin{equation}
  \varphi_{0}(r)= \frac{\beta}{M_{\rm Pl}} \int \! \frac{e^{-m \left(r-r'\right)}}{\left|r-r'\right|} \rho(r') \,r'^2 dr' \,.
\label{staticsolution}
\end{equation}

{\bf Point charge:} Treating the neutron star as a pointlike object, the mass density is
\begin{equation}
  \rho(r)=M_{1} \frac{\delta(r)}{4\pi} \,.
\end{equation}
Hence, the static solution for the scalar field simplifies to
\begin{equation}
  \varphi_{0}(r) = \left(\frac{\beta}{4\pi M_{\rm Pl}}\right)
\frac{M_{1} e^{-m r}}{r} + \varphi_\infty\,, 
\end{equation}
with $\varphi_{\infty}$ being an integration constant. 

If we now imagine a second neutron star with mass $M_{2}$, we can use the previous equation to compute the scalar force from the first neutron star that acts on the second one:
\begin{equation}
\boldsymbol{F}_{\varphi}
= \frac{\beta}{M_{\rm Pl}} \, M_{2} \boldsymbol{\nabla}\!\varphi= \frac{\beta^{2}}{4 \pi M_{\rm Pl}^2} M_{1} M_{2} \boldsymbol{\nabla}\!\left(\frac{e^{-m r}}{r}  + \varphi_\infty \right)
\,.
\label{Fphi}
\end{equation}
It is convenient to define a potential energy $V_{\varphi}$, in analogy to the gravitational potential energy: 
\begin{equation}
  V_{\varphi}= 
  - G \alpha_1 \frac{M_1 e^{-m r}}{r}
  \,,
\end{equation}
where we refer to $\alpha_i$ as the charge of the i'th neutron star. This yields a scalar force:
\begin{equation}
  \boldsymbol{F}_{\varphi} 
  = - \alpha_{2} M_{2} \boldsymbol{\nabla} V_{\varphi} 
= G \alpha_{1} \alpha_{2} M_{1} M_{2} \boldsymbol{\nabla}\!\left(\frac{e^{-m r}}{r}\right)\,.
\label{Fs}
\end{equation}
Note that in the limit where $m \rightarrow 0$, $\alpha_1=\alpha_2=1$ yields a scalar force of equal strength to gravity. If we next assume the masses and the charges of the two neutron stars to be equal, $M_{1}=M_{2}\equiv M$ and $\alpha_{1}=\alpha_{2}\equiv \alpha$, comparing Eqs.~\eqref{Fphi} and~\eqref{Fs}, we can relate the coupling $\beta$ to the charge $\alpha$ by
\begin{equation}
  \alpha^2 = 2 \beta^2 \,.
\label{alpha}
\end{equation}
Hence, in $f(R)$ gravity, where $\beta=6^{-1/2}$, the scalar charge $\alpha$ of a neutron star approximated as a pointlike object is equal to $\alpha=3^{-1/2} \simeq 0.58$.

{\bf Constant density sphere:} As a second example, we treat the neutron stars as extended objects and assume that the matter energy-momentum tensor  
$T^{\mu}_{\mu}$ is constant inside the neutron star and vanishes outside it, i.e.
\begin{equation}
T^{\mu}_{\mu} = \left\{
\begin{matrix}
-\rho_c\qquad {\rm for}\;\;\; r<R_{\rm c} \cr
\ 0 \qquad {\ \ \ \rm for}\;\;\; r>R_{\rm c}
\end{matrix}
\right. \,,
\label{rho}
\end{equation}
where $R_{\rm c}$ is the radius of the neutron star. 

The scalar field outside the star is given by~\cite{Khoury:2003rn} 
\begin{eqnarray}
\nonumber
& & \varphi_{0}(r)\simeq \left(\frac{\beta}{4\pi M_{\rm Pl}}\right)
\frac{M e^{-m_\varphi r}}{r} + \varphi_\infty 
\qquad \;\;\;\;\;\;\;\;\;\;\;\;\;\;\;\;{\rm if}\;\;\; \frac{\Delta R_{\rm c}}{R_{\rm c}} > 1\;; \\
& & \varphi_{0}(r)\simeq \left(\frac{\beta}{4\pi M_{\rm Pl}}\right)
\left(\frac{3\Delta R_{\rm c}}{R_{\rm c}}\right)
\frac{Me^{-m_\varphi r}}{r} + \varphi_\infty \qquad {\rm if}\;\;\; \frac{\Delta R_{\rm c}}{R_{\rm c}} \ll 1\,,
\label{phisummary}
\end{eqnarray}
where
\begin{equation}
\frac{\Delta R_{\rm c}}{R_{\rm c}} = \frac{\varphi_\infty-\varphi_c}{6\beta M_{\rm Pl}\Phi_{\rm c}}\,,
\label{DR}
\end{equation}
with $\Phi_{\rm c}=G M/R_{\rm c}$ being the Newtonian potential at the surface of the star. Here $\varphi_\infty$ is the asymptotic value of the field $\varphi$ as $r$ goes to infinity. Allowing for a more general potential $V(\varphi)$ for the scalar, we may expect that $\varphi_\infty$ is the minimum of the potential $V(\varphi)$ (zero for a purely massive field), while $\varphi_{\rm c}$ is the minimum of the effective potential $V_{\rm eff}(\varphi) = V(\varphi) + \beta \rho_{\rm c} \varphi / M_{\rm Pl}$. Finally, $M$ is the mass of the neutron star and $m_{\varphi}$ is the mass of the scalar field at $\varphi_\infty$.

From this result, we conclude that the charge of the neutron star is:
\begin{eqnarray}
\nonumber
& & \alpha^2  = 2\beta^2 
\qquad \;\;\;\;\;\;\;\;\;\;\;\;\;\;\;\;{\rm if}\;\;\; \frac{\Delta R_{\rm c}}{R_{\rm c}} > 1\,, \\
& & \alpha^2  = 2\beta^2 
\left(\frac{3\Delta R_{\rm c}}{R_{\rm c}}\right) 
\qquad {\rm if}\;\;\; \frac{\Delta R_{\rm c}}{R_{\rm c}} \ll 1\,.
\label{eq:chameleonscreening}
\end{eqnarray}
In the former case, we obtain the expected result for a point charge. In the latter case, the charge of the neutron star can be {\em screened} via the Chameleon screening mechanism~\cite{Khoury:2003rn}. Below, we focus on cases where the scalar charge of the neutron stars is not screened. 

\subsection{Inspiral dynamics}

In this section, we consider a binary system of two neutron stars with masses $M_1$ and $M_2$ separated by a distance $\Delta$. The inspiral dynamics of this binary can be described, to lowest order, as two Newtonian point particles. In the presence of a universally coupled massive scalar, the neutron stars carry scalar charges $\alpha_1$ and $\alpha_2$ respectively, and experience an (attractive) scalar force with magnitude
\begin{equation}
|F_{\rm s}| = \frac{G \alpha_1M_1 \alpha_2 M_2}{\Delta^2} \left(1+m\Delta \right) e^{-m\Delta} \,,
\end{equation}
where the mass of the scalar is $m$ and the range of the scalar force is characterized by the Compton wavelength $\lambda = 1/m$. The scalar force does not operate if the scalar field is very heavy, i.e. if $\lambda$ is smaller than the neutron star radius, or the scalar charge goes to zero. 

In the final stages of the inspiral phase, which is of most interest for GW observations, orbits are expected to have circularized, in which case the orbital frequency $\Omega$ is related to the separation $\Delta$ via the modified Keplerian relation
\begin{equation}\label{Kepler}
\Omega^2 = \left( \frac{G M_{\rm tot}}{\Delta^3} \right) \left[ 1 + \alpha_1 \alpha_2(1 + m\Delta) e^{-m \Delta}  \right] ~,
\end{equation}
where $M_{\rm tot} = M_1 + M_2$. Setting the binding energy to zero for infinitely separated stars, the total energy of this binary system is
\begin{equation}\label{systemenergy}
E = - \frac{G \mu M_{\rm tot}}{\Delta}  (1+ \alpha_1 \alpha_2 e^{-m \Delta})+ \frac{1}{2} \mathcal{I} \Omega^2 \,,
\end{equation}
with $\mu = M_1M_2/M_{\rm tot}$ the reduced mass, and $\mathcal{I}$ the moment of inertia for the neutron star system
\begin{align}
\mathcal{I} &= 
        \tilde{\mathcal{I}}_1 M_1 R_1^2 
        + \tilde{\mathcal{I}}_2 M_2 R_2^2
        + \mu \Delta^2\,,
\end{align}
where $R_{1,2}$ are the radii of the two neutron stars and $\tilde{\mathcal{I}}$ their dimensionless moment of inertia (e.g. $\tilde{\mathcal{I}}_{1,2} = 2/5$ when approximating neutron stars as constant density spheres). For simplicity, below we neglect the spin of the neutron stars as well as their internal structure and assume $\mathcal{I} \simeq \mu \Delta^2$. 

During the inspiral phase the orbit decays due to the emission of gravitational and scalar radiation\footnote{Rigorously speaking,
a massive scalar field does not radiate energy to infinity but since it does take energy away from the otherwise isolated object, we
would indistinctly refer to it as radiation.}. Taking the time derivative of the total energy Eq.~(\ref{systemenergy}), the energy loss rate is
\ba\label{eq:eloss}
\frac{dE}{dt} = A\left(\alpha_1,\alpha_2; m\Delta \right)~\mu \Delta^2 \Omega \frac{d\Omega}{dt}\,.
\ea
Here we have defined a dimensionless coefficient
\ba
A\left(\alpha_1,\alpha_2; m\Delta \right) = 1- \frac{4\left( 1 + \alpha_1 \alpha_2(1 + m\Delta) e^{-m \Delta}  \right)}{3 + \alpha_1 \alpha_2\left(3 + 3 m\Delta + m^2 \Delta^2\right) e^{-m \Delta} }\,,
\ea
which goes to $-1/3$ in the GR limit. 

The power emitted in gravitational waves is related to the quadrupole moment of the binary mass as
\begin{equation}\label{eq:gpower}
        \frac{dE_{\rm g}}{dt} =
                \frac{32 G \mu^2 \Delta^4 \Omega^6}{5} \propto \Omega^{10/3}\,.
\end{equation}
The gravitational waves are emitted at a frequency $f = \Omega / \pi$, e.g. twice the orbital frequency. 

The scalar radiation, however, is related to the dipole moment of the scalar charge for circular orbits (where the monopole vanishes). 
The power emitted in scalar radiation is 
\begin{equation}\label{eq:spower}
        \frac{dE_{\rm s}}{dt} =
                \frac{G \mu^2  \Delta^2 (\alpha_1 - \alpha_2)^2 \Omega^4 }{6} \propto \Omega^{8/3}\,.
\end{equation}
Hence, scalar radiation is emitted at a frequency $f = \Omega / 2\pi$, e.g. at the orbital frequency. 

If there is no other form of energy loss, we can identify 
\ba\label{eq:totalen}
\frac{dE_{\rm g}}{dt} + \frac{dE_{\rm s}}{dt} = - \frac{dE}{dt}\,,
\ea
which allows us to find the time derivative of the angular frequency
\ba
\frac{d\Omega}{dt} = - \frac{G \mu \Omega^3}{A\left(\alpha_1,\alpha_2; m\Delta \right)} \left( \frac{32}{5} \Delta^2 \Omega^2 + C \right)
\label{domega}\ea
and then combining Eqs.~(\ref{eq:gpower}), (\ref{domega}), and \eqref{eq:totalen} we get
\ba\label{dEdOmega}
\frac{dE_{\rm g}}{d\Omega} =  \mu \Delta^4 \Omega^3  \frac{A\left(\alpha_1,\alpha_2; m\Delta \right)}{G^{2/3} M_{\rm tot}^{2/3} B^{2/3} \Omega^{2/3} + \frac{5}{32}C},
\ea
where we defined
\ba\label{B}
B = 1 + \alpha_1 \alpha_2(1 + m\Delta) e^{-m \Delta} \quad \text{and} \quad C = \frac{1}{6} (\alpha_1 - \alpha_2)^2\,.
\ea
Note that the power emitted in gravitational radiation ($\propto \Omega^{10/3}$) ramps up faster as the orbit decays, than the power emitted in scalar radiation ($\propto \Omega^{8/3}$). Therefore, we expect scalar radiation to dominate at low frequencies, or equivalently at large separations. If the separation is larger than the one where the power in scalar radiation equals that in gravitational radiation, $\Delta > \Delta_{\rm eq}$, one may therefore conclude that the scalar radiation can be an important factor to consider in the orbital decay. This is true for a massless scalar field.

When the scalar has a mass, we must alter the story above somewhat. It becomes possible for scalar radiation to remain gravitationally bound to the orbital system, and therefore energy does not meaningfully leave the system. As noted above, the frequency of the emitted scalar radiation is equal to the orbital frequency of the binary. Therefore, there will be a critical separation for the binary, below which radiation has sufficient energy to escape the gravitational potential of the binary, and hence carry energy away. We can estimate this critical separation, $\Delta_{\rm crit}$, by requiring the momentum of the scalar waves to be larger than the escape momentum, for a particle of mass $m$, from the binary. The momentum of scalar waves is
\begin{equation}
p_{\rm m} = \sqrt{\Omega^2 - m^2}\,,
\end{equation}
where we can approximate $\Omega \simeq \sqrt{G M_{\rm tot}/\Delta^3}$. The minimal momentum for a particle of mass $m$ to escape from the binary is
\begin{equation}
p_{\rm esc} \simeq m \sqrt{\frac{2 G M_{\rm tot}}{\Delta}}.
\end{equation}
For $\Delta \gg 2 G M_{\rm tot}$
\begin{equation}
\Delta_{\rm crit} \simeq (2 G M_{\rm tot} / m^2)^{1/3},
\end{equation}
which is between the Schwarzschild radius associated with the binary and the Compton wavelength of the scalar. When the mass of the scalar is small, this critical value is smaller than the Compton wavelength of the scalar. When the mass of the scalar is instead large, $\Delta_{\rm crit}$ can be larger than the Compton wavelength of the scalar, but still smaller than the Schwarzschild radius. So, we can summarize by saying that scalar radiation can carry away energy only when: 1) the Compton wavelength of the scalar is larger than the Schwarzschild radius associated with the binary and 2) the members of the binary system are a bit closer than the Compton wavelength of the scalar. Note that this is roughly what you expect, since the scalar force between the members of the binary only turns on when they are closer than a Compton wavelength apart.

As we mentioned above, scalar radiation can carry energy away from the system only when $\Delta < \Delta_{\rm crit}$ and the power in scalar radiation dominates that in gravitational radiation when $\Delta > \Delta_{\rm eq}$. Therefore, scalar radiation will contribute significantly to the orbital dynamics when $\Delta_{\rm eq} < \Delta_{\rm crit}$. Since $m \,\Delta_{\rm crit} \sim \mathcal{O}(1)$, we conclude that $m \,\Delta_{\rm eq}$ needs to be smaller than one, hence using Eqs.~\eqref{eq:gpower} and~\eqref{eq:spower}, $\Delta_{\rm eq}$ can be approximated as
\begin{equation}
\Delta_{\rm eq} \simeq \frac{192\, G M_{\rm tot}}{5} \frac{1+\alpha_{1} \alpha_{2}}{\left(\alpha_{1} -\alpha_{2}\right)^2} \,.
\end{equation}
Here, we have used that $\Omega \simeq \sqrt{\left(G M_{\rm tot}/\Delta^3\right) (1+\alpha_{1} \alpha_{2})}$ (compare with Eq.~\eqref{Kepler}).
Consequently, the criterion $\Delta_{\rm eq} < \Delta_{\rm crit}$  implies that we must have a Compton wavelength $\lambda$ greater than a certain value, as
\ba
\lambda > 2 \,G M_{\rm tot} \left(\frac{96}{5} \frac{1+\alpha_{1} \alpha_{2}}{(\alpha_1-\alpha_2)^2} \right)^{3/2} \,.
\ea
Unless the magnitude of the charge dipole is $\mathcal{O}(1)$, the Compton wavelength of the scalar must be large compared to the scales characterizing the final stages of inspiral (e.g. $2GM_{\rm tot}$). This implies that we can safely neglect scalar radiation for scenarios where the Compton wavelength is of order the size of individual neutron stars. 

For this reason, we will simply neglect the effects caused by the scalar radiation by setting $C=0$ in the following, in which case Eq.~\eqref{dEdOmega} simplifies to
\ba
\frac{dE_{\rm g}}{d\Omega} =  (-3A B^{2/3}) (-\frac13 G^{2/3}M_{\rm tot}^{2/3}\mu \Omega^{-1/3}),
\ea
where $\Omega \simeq \sqrt{G M_{\rm tot}/\Delta^3}$.
When the separation of two stars is much larger than the Compton wavelength of the scalar field, i.e. $m\Delta > 1$, we have $A \simeq -1/3$ and $B \simeq1$, and we recover the GR result

\ba
\frac{dE_{\rm g}}{d\Omega} \simeq -\frac13 G^{2/3} \mu M_{\rm tot}^{2/3} \Omega^{-1/3}~.
\ea
When the separation is much smaller than the Compton wavelength of scalar field, we have
$A \simeq -1/3$ and $B \simeq1+\alpha_1\alpha_2$, and we get
\ba \label{dENGR}
\frac{dE_{\rm g}}{d\Omega} \simeq -\frac13 G^{2/3}\mu M_{\rm tot}^{2/3} (1+\alpha_1\alpha_2)^{-2/3}  \Omega^{-1/3}~,
 \ea
in which case the power is enhanced by a factor of $(1+\alpha_1\alpha_2)^{-2/3}$. 
It is also instructive to calculate the total energy emitted during the inspiral phase. In the absence of other forms of dissipation, we can estimate the total energy emitted in radiation (gravitational and scalar) to be
\ba
E_{{\rm insp}} = \left( \frac{G \mu M_{\rm tot}}{\Delta}  (1+ \alpha_1 \alpha_2 e^{-m \Delta}) - \frac{1}{2} \mathcal{I} \Omega^2 \right)_{\Delta = R_1+R_2} .
\ea
To get an idea for the potential boost in radiated energy, consider the case of identical neutron stars ($M=M_1=M_2$ and $R=R_1=R_2$). For equal scalar charge, we can neglect scalar radiation (as described below, a scalar charge dipole is necessary to efficiently emit scalar radiation). In this scenario, assuming that $\lambda \gg R_1+R_2$, we have 
\ba
E_{\rm insp} = \frac{GM^2}{4R} \left(1 - \tilde{\mathcal{I}} \right) (1+\alpha^2) = (1+\alpha^2) E_{\rm insp, GR} \,.
\ea 
In the presence of the scalar force, the total energy emitted in gravitational radiation during the inspiral phase is boosted by a factor of $1+\alpha^2$ as compared to GR. For $\alpha^2 \simeq 1/3$, a choice relevant to our discussion below, the total energy emitted during inspiral can be up to $\sim 33\%$ higher in the presence of the scalar force.

\noindent

\subsection{Inspiral waveform}

Given the inspiral dynamics, we now examine the gravitational waveform in the presence of a finite-range scalar force. The measured strain $h(t)$ in a gravitational wave detector is~\cite{Flanagan:1997sx,Allen:2005fk}
\ba\label{waveform}
h(t) = \frac{4 Q}{D_L} G \mu\, \Omega^2 \Delta^2 \cos\left(\int 2\pi f dt\right),
\ea
where $D_L$ is the luminosity distance to the source and $Q$ encodes the detector response as a function of the angular position and orientation of the binary. In the following, we set $Q=1$ for convenience (equivalent to perfect response to either of the two GW polarizations). In addition, we neglect the cosmological red-shifting of the observed frequency of gravitational radiation (motivated by the limited horizon for neutron star binary mergers with LIGO). In the presence of a scalar force, the GW frequency evolves as
\ba\label{dOmegadt}
\frac{df}{dt} =\frac{B^{2/3}}{-3A} \frac{96}{5} \pi^{8/3}G^{5/3} {\cal M}^{5/3} f^{11/3}\,,
\ea
where ${\cal M} = \mu ^{3/5} M_{\text{tot}}^{2/5}$ is the chirp mass.

\begin{figure}[t]
\centering 
\includegraphics[height=0.3\textwidth]{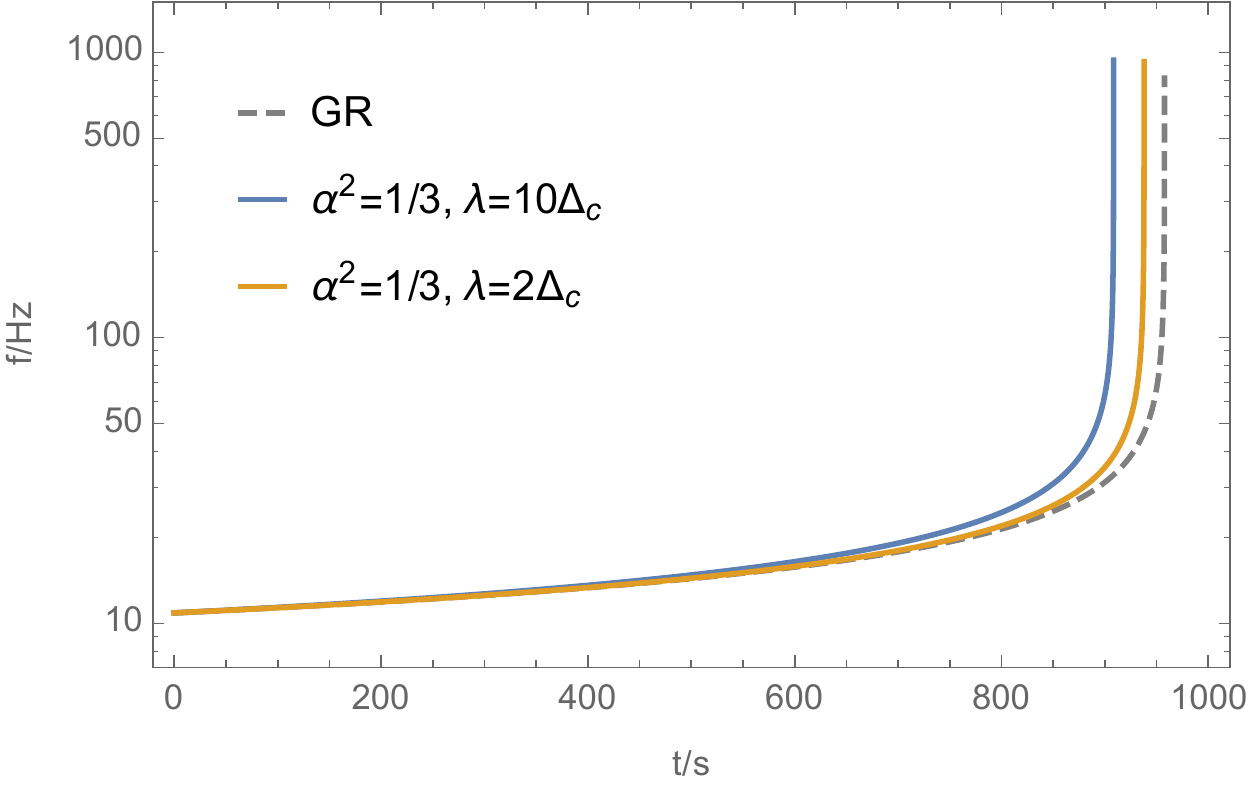}
\includegraphics[height=0.3\textwidth]{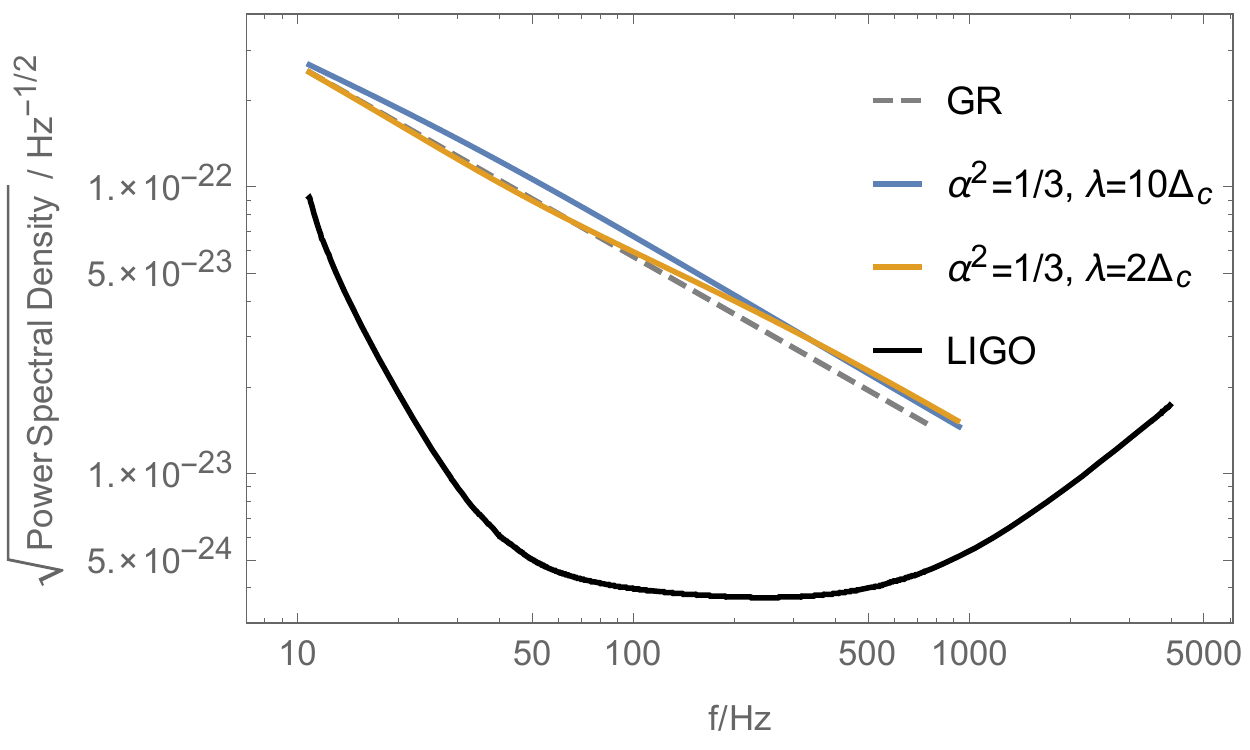}
\caption{\label{fig:wave} GW signals from a neutron star binary with $M_1 = M_2 = 1.25M_{\rm \odot}$ and $D_L=135 \ {\rm Mpc}$ in the Newtonian approximation. We assume that the compaction of the neutron stars is $0.1$, and that the merger happens (and therefore the plots are cut off) when the two neutron stars contact. Also, the geometrical factor has been taken to be perfect response. The gray dashed line are signals predicted by GR. The blue and orange lines are signals with the presence of a scalar force of different  Compton wavelength. The left panel shows the time dependence of the frequency. The right panel shows the square root of the power spectral density against the LIGO sensitivity curve (the line in black). }
\end{figure}

In the left panel of Fig.~\ref{fig:wave}, we plot $f(t)$ for GR and two merger scenarios with a scalar force (with $\alpha^2=1/3$ for both and two choices of the Compton wavelength $\lambda = 2 \Delta_{\rm c}, 10 \Delta_{\rm c}$). We track the evolution until the stars come into contact at a separation $\Delta_{\rm c}$, here assumed to be $\Delta_{\rm c} = G M_{\rm tot}/C$ with identical compaction $C \equiv GM/R = 0.1$ for both merging stars. Generally speaking, the presence of the scalar force accelerates the merger and produces a higher pitched chirp (higher peak frequency). The merger takes place earlier and the peak frequency is higher, as the Compton wavelength of the scalar increases.

We will also need the Fourier transform of the time-domain waveform, given by
\ba
\h(f) \equiv \int_{-\infty}^{\infty} e^{2\pi ift}h(t) dt.
\ea
For a strain of a form $h(t) = h_0(t) \cos\phi(t)$, where $d\ln h_0/dt \ll d\phi/dt$ and $d^2\phi/dt^2 \ll (d\phi/dt)^2$, $\h(f)$ can be computed using the stationary phase approximation, 
\ba
\h(f) \simeq H(f) e^{i \Psi(f)},
\ea
where 
\ba
H(f) = \frac12 h_0(t) \left(\frac{df}{dt}\right)^{-1/2},
\ea
and
\ba\label{eq:fullpsi}
\Psi(f) =  2\pi f t-\phi(f) -\frac{\pi}{4}.
\ea
In the above two equations,  $t$ should be thought as a function of $f$ and is defined as the time at which $d\phi/dt = 2\pi f$. Using Eqs.~(\ref{Kepler}), (\ref{waveform}) and (\ref{dOmegadt}), we  find the amplitude of $\h(f)$ in the presence of the scalar force, given by
\ba
H(f) = \left(-3A(f)\right)^{1/2} B(f)^{1/3} \sqrt{\frac{5}{24}} \frac{G^{5/6} {\cal M}^{5/6}}{ \pi^{2/3} D_L} f^{-7/6}\,.
\ea
An analytic expression for the phase cannot be obtained in the general case. However, in the limit where $m \rightarrow 0$, we get
\ba
\Psi(f) &=& 2\pi t_c f - \frac{\pi}{4} +\frac{3}{4 (1+\alpha^2)^{2/3}} \left(8 \pi G {\cal M} f \right)^{-5/3} -\phi_{\rm c}\,.
\ea
Here, $t_c$ is the time at which the separation (formally) goes to zero in the Newtonian limit of coalescing point particles, and $\phi_c$ is the phase at this time. More generally, we integrate $d\phi/dt = 2\pi f$ to obtain $\Psi$ from Eq.~\eqref{eq:fullpsi}. We set the integration constant by choosing a value for $\Psi$ at a particular frequency $f_{\rm eq}$.

\begin{figure}[t]
\centering 
\includegraphics[height=0.3\textwidth]{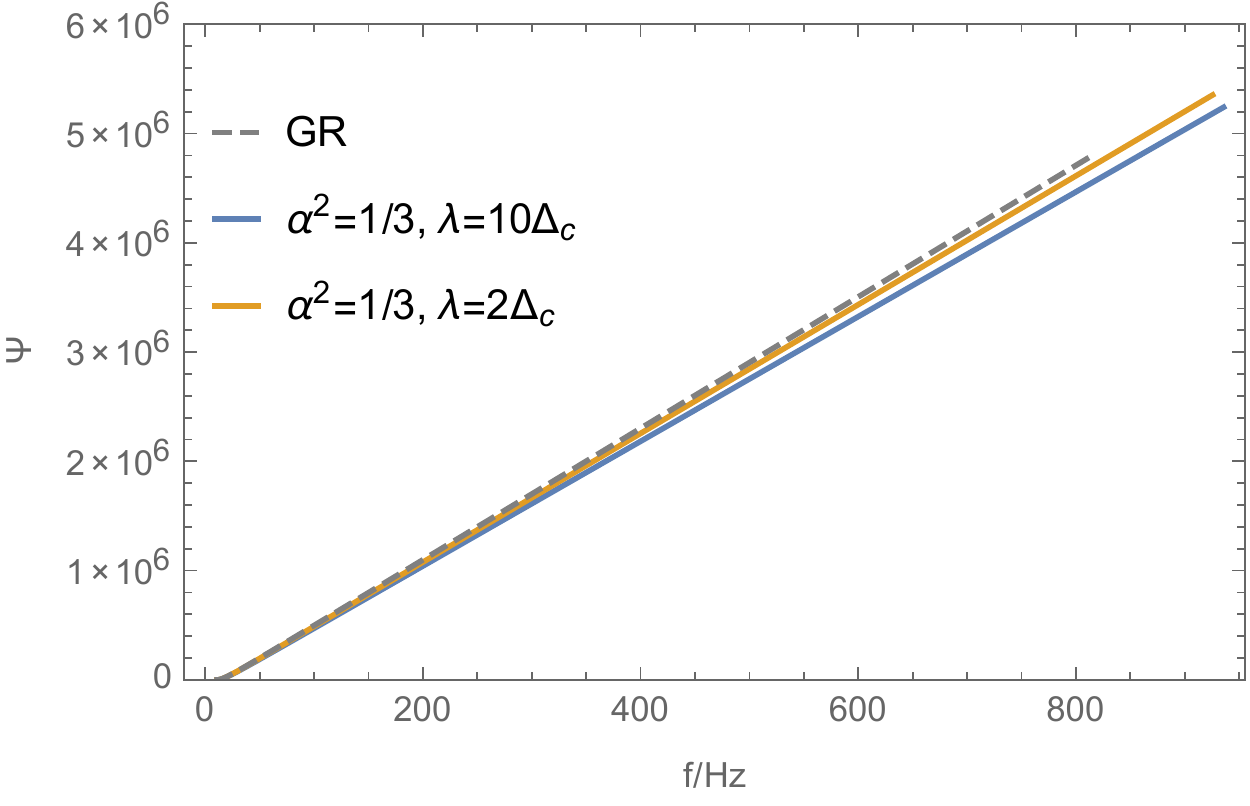}
\caption{\label{fig:phase} Phases of GW signals predicted by GR (the gray dashed line) and theories in the presence of a scalar force (the blue and orange lines). The parameters are the same as those in Fig.~\ref{fig:wave}. To show the differences, the phases are calibrated so that they are identical at the low frequency boundary of the LIGO sensitivity window.}
\end{figure}

In the right panel of Fig.~\ref{fig:wave}, we show the square root of the power spectral density $\sqrt{S_h(f)} = 2 f^{1/2} H$ for GR and the two merger scenarios described above plotted against the projected sensitivity of advanced LIGO~\cite{Aasi:2013wya}. We have assumed that the binary is located at $D_L=135 \ {\rm Mpc}$ in this example. For Compton wavelengths significantly larger than the separation at merger, $\Delta_{\rm c}$, the scalar force has a significant effect on the amplitude within the frequency window probed by LIGO. The phase $\Psi(f)$ also plays an important role in distinguishing waveforms. We show the phase within the LIGO sensitivity window in Fig.~\ref{fig:phase}, where it can be seen that the difference in phase is significant. In this figure we set the phases equal at $f_{\rm eq}=10$ Hz by choosing a set of integration constants appropriately. 

\subsection{Forecasted constraints from the inspiral waveform}\label{sec:forecast}

Given a high signal-to-noise detection of a neutron star merger, it is possible to use the measured inspiral waveform to derive constraints on the product of charges $\alpha_1\alpha_2$ and Compton wavelength $\lambda$ in the general model described above. Given a measured signal $s(t,\boldsymbol{\bar{\theta}}) = n(t) + \bar{h}(t,\boldsymbol{\bar{\theta}})$ consisting of a noise realization $n(t)$ and merger waveform $\bar{h}(t,\boldsymbol{\bar{\theta}})$ depending on the ``true" parameters $\boldsymbol{\bar{\theta}}$, as well as a set of template wave-forms $g(t,\boldsymbol{\theta})$ depending on a set of candidate parameters $\boldsymbol{\theta}$ (the
best fit parameters in the presence of some realization of the noise), the likelihood function is given by
\begin{equation}
\mathcal{L} (s | \boldsymbol{\theta}) = \mathcal{N} \exp\left[ -\frac{1}{2} \left(\,s - g \,|\,s-g\,\right) \right]\,, 
\end{equation}
where $\mathcal{N}$ is a normalization factor \cite{Finn:1992wt}. Given two signals $h(t)$ and $g(t)$, the inner product $\left(\,h\,|\,g\,\right)$ on the vector space of signals is defined as
\ba
\left(\,h\,|\,g\,\right) = 2\int_0^{\infty} \frac{\h^*(f) \tilde{g}(f) + \h(f)\tilde{g}^*(f)}{S_n(f)}df\,,
\ea
where $S_n(f)$ is the detector noise spectral density and $\h, \tilde{g}$ are the Fourier transforms of $h, g$. The inner product is defined so that the probability of having a noise realization $n_0(t)$ is $p(n=n_0) \propto \exp[-(n_0|n_0)/2]$. We then marginalize the logarithm of the likelihood over many noise realizations (e.g.~\cite{Cutler:1994ys})
\ba\label{eq:dchisq}
\langle \Delta \chi^2 (\boldsymbol{\theta}) \rangle &\equiv& 2 \langle \log\left[ \mathcal{L} (s | \boldsymbol{\theta}) / \mathcal{L} (s | \bar{\boldsymbol{\theta}}) \right] \rangle \nonumber \\
&=&  \left(\,\bar{h} - g \,|\, \bar{h}-g\,\right) \nonumber \\
&=& 4 \int_0^\infty \frac{df}{S_n(f)} \left( H(f,\bar{\boldsymbol{\theta}})^2 + H(f,\boldsymbol{\theta})^2 - 2H(f,\bar{\boldsymbol{\theta}}) H(f,\boldsymbol{\theta}) \cos \left[ \Psi (f, \boldsymbol{\theta}) - \Psi (f, \bar{\boldsymbol{\theta}}) \right] \right)\,,
\ea
where $\mathcal{L} (s | \bar{\boldsymbol{\theta}})$ is the likelihood evaluated at $g=\bar{h}$ with $H$ and $\Psi$ the amplitude and phase of the waveform in the stationary phase approximation.

To give an idea of the constraints on the parameters in the sector of the scalar force, we consider a parameter space including  $\boldsymbol{\theta} = \{\alpha^2 \equiv \alpha_1 \alpha_2, \lambda, {\cal A}, {\cal M}, M_{\text{tot}}, t_c, \phi_c \}$, where ${\cal A} \equiv \sqrt{\frac{5}{24}} \frac{G^{5/6} {\cal M}^{5/6}}{ \pi^{2/3} D_L}$. Note that since the dynamics of the binary depend only on the product $\alpha_1\alpha_2$ it is not possible to derive constraints on the two charges individually, at least in the Newtonian limit we are working in (and in the absence of the effects of scalar radiation, which could in principle be used to break this degeneracy as it depends on the charge dipole; see Eq.~(\ref{eq:spower})).We consider two fiducial scenarios, focusing on $\alpha^2$ and $\lambda$. In the first one, the binary evolves according to GR (e.g. the case of a binary merger in the absence of a scalar force), $\bar{\boldsymbol{\theta}} = \{\alpha^2 \rightarrow 0, \lambda \rightarrow 0,\,... \}$ and we assume that $M_1= M_2=1.25 M_{\odot}$, and $D_L=40 \ {\rm Mpc}$. 
This provides an idea of the noise-limited constraints that could be obtained by Advanced LIGO for a nearby event. In the second case, the signal contains a binary influenced by a scalar force with $\bar{\boldsymbol{\theta}} = \{\alpha^2 =1/3, \lambda = 7 G M_{\odot},\, ... \}$ and the other parameters the same as the first scenario. For both scenarios, we perform Markov chain Monte Carlo (MCMC) sampling using the {\em emcee} package~\cite{2013PASP..125..306F} to sample the likelihood function on the full 7-dimensional parameter space. In addition, we use both the observed noise curve during the first run of 
LIGO ("O1")~\cite{Martynov:2016fzi}, and the forecasted noise curve for advanced LIGO at design sensitivity ("Design") based 
on the Zero Det, High Power scenario~\cite{LIGOnoise}. 

\begin{figure}[t]
\centering 
\includegraphics[height=0.3\textwidth]{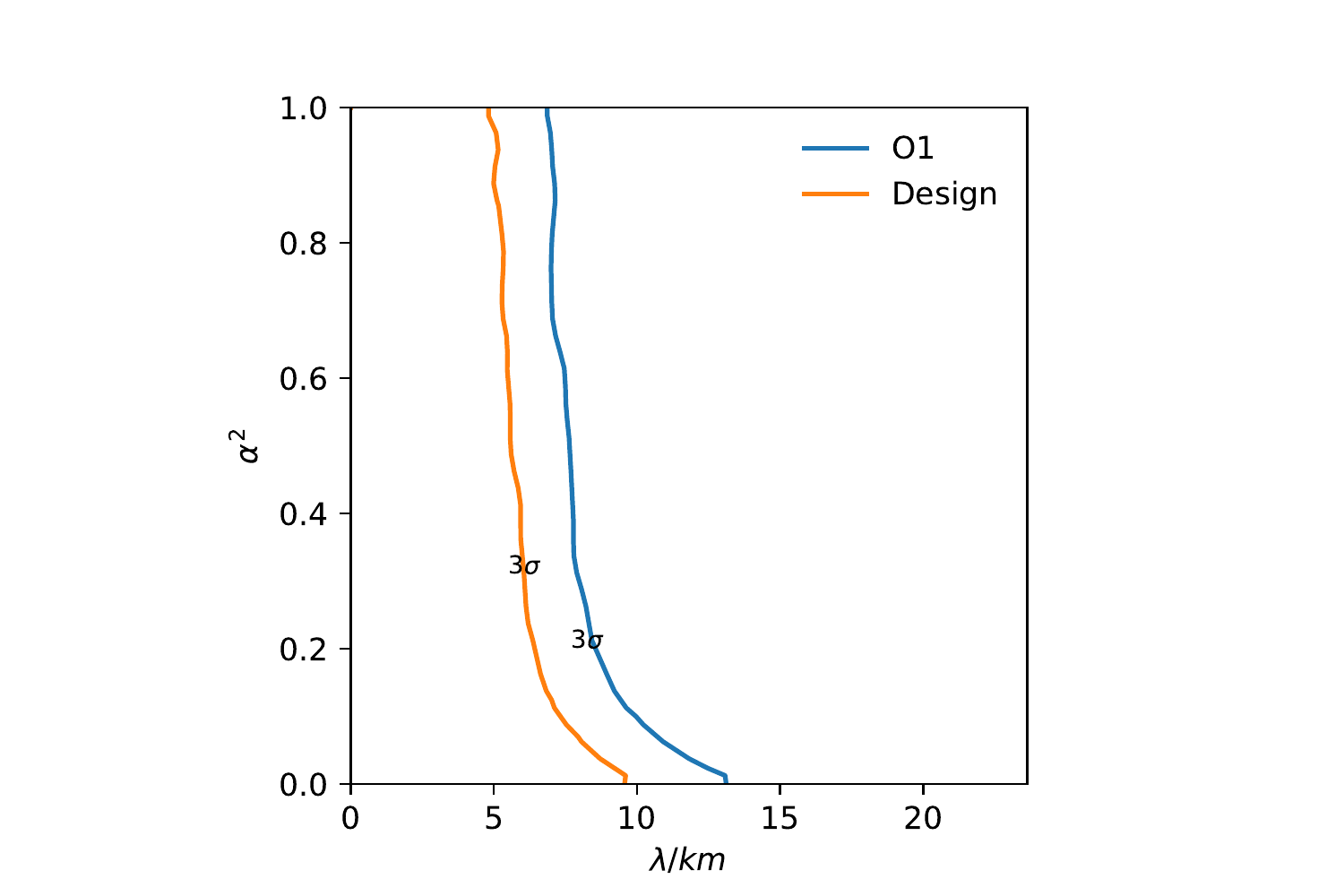}
\includegraphics[height=0.3\textwidth]{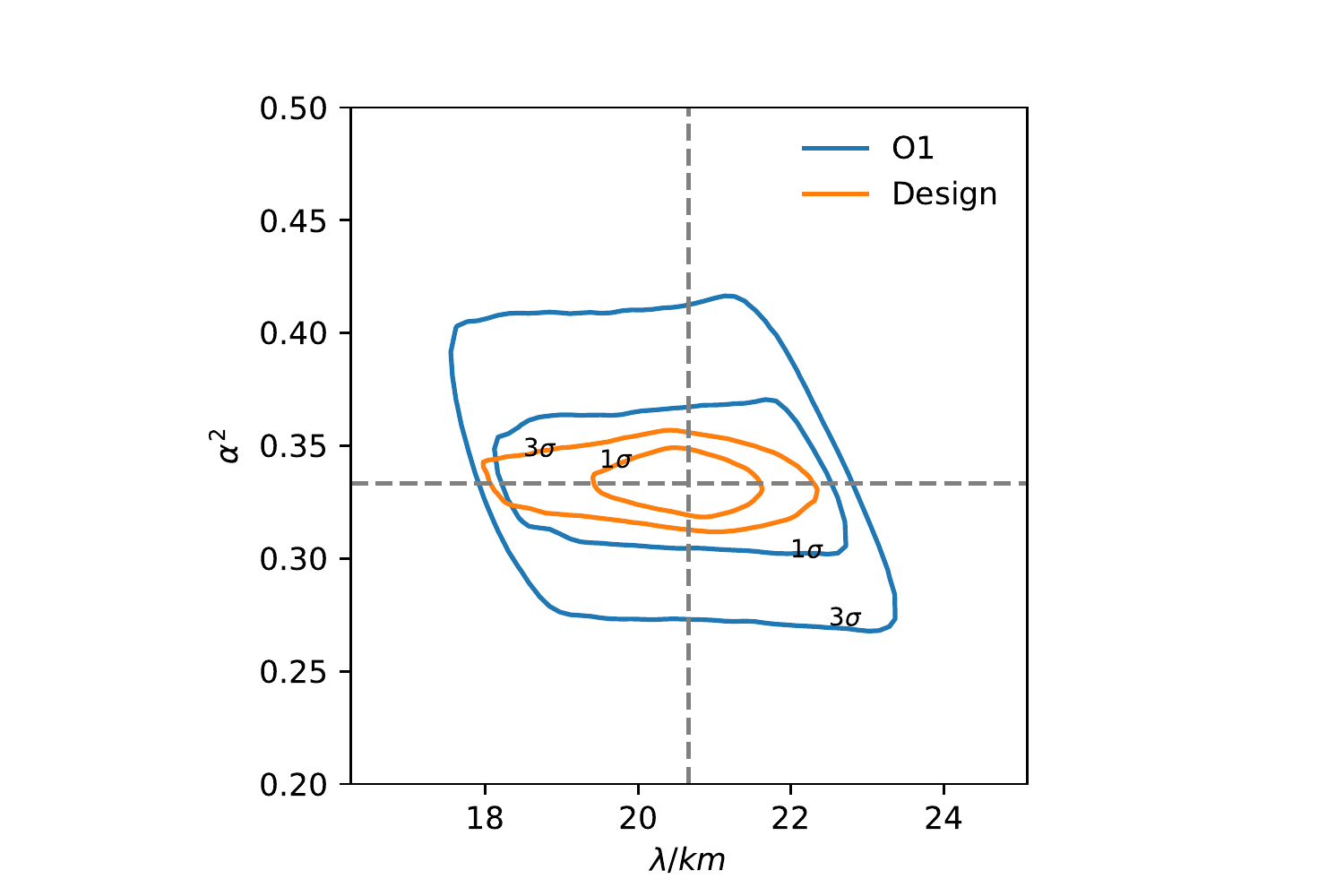}
\caption{\label{fig:deltachi} The marginalized constraints at one and three $\sigma$ in the $\alpha^2$ - $\lambda$ plane with $\bar{\boldsymbol{\theta}} = \{\alpha^2 \rightarrow 0, \lambda \rightarrow 0, \, ... \}$ (left panel) and $\bar{\boldsymbol{\theta}} = \{\alpha^2 =1/3, \lambda =7 GM_{\odot}, \, ...\}$ (right panel). The other parameters are fixed by choosing $M_1= M_2=1.25 M_{\odot}$, and $D_L=40 \ {\rm Mpc}$ in both scenarios. The errors are calculated with both LIGO's O1 noise curve (in blue) and design noise curve (in orange).}
\end{figure}

The result is shown in Fig.~\ref{fig:deltachi}. In the left panel, we show the noise-limited 3-$\sigma$ constraints in the absence of a scalar force. For this fiducial merger, constraints on the Compton wavelength of order $\lambda \alt \mathcal{O} (5-10 \ {\rm km})$ can be placed over a wide variety of scalar charges. In addition, it can be seen that there is little improvement in the constraint when comparing O1 and Design sensitivity, and so already with one detection at current sensitivity LIGO should be able to place stringent bounds. In the right panel, we show the one and three sigma contours obtained from the case with a non-zero scalar charge and compton wavelength. Here, there is a more dramatic improvement between O1 and Design sensitivity. At design sensitivity, and for this fiducial event, LIGO would be able to determine the charge and compton wavelength of the scalar at the $\sim 10 \%$ level.

In conclusion, the inspiral phase of neutron star mergers appears to be a powerful laboratory for constraining scalar forces weaker than gravity with ranges down to the $\lambda \sim \mathcal{O} (5-10 \ {\rm km})$ scale, or providing the basis for a possible detection if such scalar fifth forces are realized in nature. This brief analysis motivates a more systematic study of the possible constraints given a realistic source population and using the full parameter space, which we defer to future work.

\section{Metric $f(R)$ Gravity}
\label{Sec:metricfR}

One of the simplest extensions of Einstein's theory of general relativity  is $f(R)$ modified gravity (see, e.g.,~\cite{Sotiriou:2008rp,DeFelice:2010aj,Capozziello:2011et} for reviews). In $f(R)$ modified gravity, the Einstein-Hilbert action of GR is generalized by replacing the Ricci scalar $R$ with an arbitrary function $f(R)$. Hence, the action for $f(R)$ gravity reads~\cite{Sotiriou:2008rp} 
\begin{equation}
S=\frac{1}{16 \pi G} \int\! d^4 x \,\sqrt{-g}\,f(R)+
S_{\rm M}[g_{\mu \nu},\psi] \,,
\label{action}
\end{equation} 
where we set 
$\hbar = c = 1$. Moreover, we defined $g \equiv \rm{det}(g_{\mu \nu})$ as the determinant of the metric tensor $g_{\mu \nu}$,
and 
$S_{M}$ denotes the action of the matter fields, which are collectively denoted by $\psi$. To avoid tachyonic instabilities and the appearance of ghosts (negative kinetic energy states), viable $f(R)$ theories require monotonically growing and convex functions $f(R)$ such that
\begin{equation}
  \frac{df(R)}{dR}>0 \,, \quad \frac{d^2f(R)}{dR^2}\ge 0 \,.
\label{ineqf}\end{equation}
Note that, in what follows, we will focus on metric $f(R)$ gravity where the field equations are obtained by varying the action Eq.~\eqref{action} with respect to the metric tensor. We will not discuss Palatini $f(R)$ gravity~\cite{Sotiriou:2008rp,Olmo:2011uz} here.

Dynamically $f(R)$ theories are equivalent to a specific class of scalar-tensor theories~\cite{Teyssandier:1983zz,Barrow:1988xi,Barrow:1988xh,Wands:1993uu}. 
Depending on which quantity is identified with the scalar field, different formulations of this dynamical equivalence exist (see \cite{Berti:2015itd}). A common choice is to transform the $f(R)$
action~\eqref{action} into a Brans-Dicke theory \cite{Brans:1961sx,Dicke:1961gz} (with $\omega=0$) in the Jordan frame, 
\begin{equation}
 S^{\rm J}=\frac{1}{16 \pi G} \int\! d^4 x\,\sqrt{-g} \, \Bigl(\phi R - U(\phi) \Bigr) +S^{{\rm J},\, {\rm M}}[g_{\mu \nu},\psi] \,, 
\label{action_J}
\end{equation}
where the scalar field $\phi$ and the Brans-Dicke potential $U(\phi)$ are defined by
\begin{equation}
  \phi \equiv \frac{df(R)}{dR} \,, 
  \quad 
  U(\phi) \equiv R \,\frac{df(R)}{dR} - f(R) \,.
\label{U}
\end{equation}
In many cases it is useful to study $f(R)$ gravity as a scalar-tensor theory in the Einstein frame. One can recast the action~\eqref{action_J} as an action in the Einstein frame by performing a conformal transformation and redefining the scalar field~\cite{Teyssandier:1983zz,Sotiriou:2006hs}

\begin{equation}
  g^{\rm E}_{\mu\nu} \equiv \phi(\varphi)\, g_{\mu\nu} \,,
  \quad
  \phi(\varphi) \equiv e^{2 \beta \frac{\varphi}{M_{\rm Pl}}}
  \,,
  \label{trafo}
\end{equation}
with $\beta = 1/ \sqrt{6}$. Thus, the action in the Einstein frame for the scalar field $\varphi$ reads
\begin{equation} 
 S^{{\rm E}}=\int\! d^4 x\,\sqrt{-g^{\rm E}} \left(\frac{M_{\rm Pl}^{2}}{2} R^{\rm E} - \frac{1}{2} \partial_{\mu}\varphi \, \partial^{\mu} \varphi - V(\varphi) \right) +S^{{\rm E},\, {\rm M}}[g^{{\rm E}}_{\mu \nu} \, \phi(\varphi)^{-1},\psi] \,, 
\label{action_E}
\end{equation}
where the scalar potential is 
\begin{equation}
  V(\varphi) \equiv \frac{U\bigl(\phi(\varphi)\bigr)}{\phi(\varphi)^{2}}  
  \,.
\label{V}
\end{equation} 
(compare definitions \eqref{U} and \eqref{trafo}). Notice that, in the presence of a non-trivial potential $V(\varphi)$, the scalar field $\varphi$ is massive; its mass arises from the potential as
\begin{equation}
  m^2\equiv \left. \frac{d^2 V(\varphi)}{d\varphi^{2}} \right|_{\varphi=\varphi_{\rm min}} \,,
  \label{m}
\end{equation} 
with $\varphi_{\rm min}$ being the field value at the minimum of the potential.


\subsection{$R^{2}$ gravity}\label{sec:r2gravity}

As a case-study, we specialize to $R^{2}$ gravity, a particular model of $f(R)$ gravity with a single free parameter. In $R^{2}$ gravity, the function $f(R)$ in the Einstein-Hilbert action \eqref{action} is set to \cite{Staykov:2014mwa}
\begin{equation}
  f(R)= R + a_{2} R^2 \,, 
\label{inequ}
\end{equation}
where $a_{2}$ is a free parameter of dimension $\left[a_{2}\right]=\left[R\right]^{-1}$. Note that the second inequality in Eq.~\eqref{ineqf} requires $a_{2} \geq 0$. 

According to Eq.~\eqref{U}, the scalar field $\phi$ and the Brans-Dicke potential $U(\phi)$ for $R^2$ gravity are given by
\begin{equation}
  \phi = 1+ 2 \, a_{2} R \,, 
  \quad 
  U(\phi) = \frac{1}{4a_{2}} \left(\phi-1\right)^2 \,.
\label{UR2}
\end{equation}
Consequently, the scalar potential $V(\varphi)$ in Eq.~\eqref{V} equals
\begin{equation}
  V(\varphi) = \frac{{M_{\rm Pl}^{2}}}{8a_{2}}
\left(1-e^{- \sqrt{\frac{2}{3}} \frac{\varphi}{M_{\rm Pl}}}\right)^2
  \,.
\label{VR2}
\end{equation} 
This allows us in turn to determine from Eq.~\eqref{m} the mass $m_{\varphi}$ of the scalar field about the vacuum at $\varphi=0$  as
\begin{equation}
  m_{\varphi} 
  = \sqrt{\frac{1}{6 a_{2}}} \,.
  \label{mR2}
\end{equation} 
Deviations from a quadratic potential about the vacuum become important for field excursions of order $\Delta\phi \agt \Delta \phi_{\rm NL}$ with $\Delta \phi_{\rm NL} \equiv \sqrt{3/2} M_{\rm Pl} \simeq 0.244 ~G^{-1/2}$. 

\subsection{Experimental constraints on $a_{2}$ 
\label{Constraints}}

Several experimental bounds which constrain the free parameter $a_{2}$ of $R^2$ gravity exist \cite{Naf:2010zy}. For example, the laboratory bound from the E\"{o}t-Wash experiment is $a_{2} \unit[\lesssim 10^{-10}]{m^{2}}$. This effectively rules out the parameter range relevant here, unless there is some novel screening mechanism that operates on the Earth but not for other astrophysical systems. Nevertheless, it is important to consider independent constraints. The tightest existing space-based constraint, coming from the satellite mission Gravity Probe B, requires $a_{2} \lesssim \unit[5 \times 10^{11}]{m^{2}}$. Measurements of the precession of the pulsar B in the PSR J0737-3039 system impose instead a less stringent limit of $a_{2} \lesssim \unit[2.3 \times 10^{15}]{m^{2}}$. 

The first neutron star mergers observed by LIGO have the potential to impose far more stringent bounds on $a_2$ than the existing astrophysical constraints. Considering only the inspiral, and assuming that $\alpha^2=1/3$ as expected from the discussion above, the predictions from $R^2$ gravity trace out a horizontal line in the left panel of Fig.~\ref{fig:deltachi}. For the binary considered in this plot, we have $\lambda \alt 5 \ {\rm km}$, and we find that it would be possible to constrain $a_2 \alt 4 \times 10^6 \ {\rm m}^2$ at $3 \sigma$. This represents an improvement of 5 orders of magnitude over the constraint imposed by Gravity Probe B.

\subsection{Equations of motion}\label{sec:eoms}

In order to derive the equation of motion for the scalar field $\varphi$, we will work in the Einstein frame. 
 By rewriting the action in the Einstein-frame, $S^{\rm E}$, in Eq.~\eqref{action_E} as a space-time integral over the corresponding Lagrangian, 
we see that this Lagrangian consists of a $\varphi$-dependent as well as a $\varphi$-independent part. From the $\varphi$-dependent part,
\begin{equation}
 \mathcal{L}^{{\rm E},\, \varphi}=- \sqrt{-g^{\rm E}} \left(\frac{1}{2} \partial_{\mu}\varphi \, \partial^{\mu} \varphi + V(\varphi) \right) + \mathcal{L}^{{\rm E},\, {\rm M}}[g^{\rm E}_{\mu \nu} \, \phi(\varphi)^{-1},\psi]  \,,
\label{Lagrangian}
\end{equation}
we can derive the equation of motion for the scalar field $\varphi$. For this purpose, we first vary $\mathcal{L}^{{\rm E},\, \varphi}$ with respect to the metric tensor $g_{{\rm E}}^{\mu \nu}$ to obtain the energy-momentum tensor $T^{{\rm E},\, \varphi}_{\mu \nu}$,
\begin{equation}
\frac{2}{\sqrt{-g}} \frac{\delta \mathcal{L}^{{\rm E},\, \varphi}}{\delta g_{\rm E}^{\mu \nu}} 
\equiv \tilde{T}^{{\rm E},\ \varphi}_{\mu \nu}  
\,.
\end{equation}
The latter is the sum of the scalar-field and matter energy-momentum tensors, $
\tilde{T}^{{\rm E},\, \varphi }_{\mu \nu} = T^{{\rm E},\, \varphi}_{\mu \nu} + T^{{\rm E},\, {\rm M}}_{\mu \nu}$,
which arise as (see also ~\cite{Tsujikawa:2008uc,Tsujikawa:2009yf}) 
\begin{equation}
    T^{{\rm E},\, \varphi}_{\mu \nu} = - g_{\mu \nu} \left(\frac{1}{2} \partial_{\alpha}\varphi \, \partial^{\alpha} \varphi + V(\varphi) \right) + \partial_{\mu}\varphi \, \partial^{\mu} \varphi \,, \quad
T^{\rm E}_{\mu \nu} = \frac{2}{\sqrt{-g}} \frac{\delta \mathcal{L}^{{\rm E},\, {\rm M}}}{\delta g_{\mu \nu}} \,.
\label{TM}
\end{equation}
Thereby, $T^{\rm E}_{\mu \nu}$ is related to the physical energy-momentum tensor $T_{\mu \nu}$ in the Jordan frame by the conformal transformation \eqref{trafo},
\begin{equation}
  T^{E}_{\mu \nu} = \phi(\varphi)^{-2} \,T_{\mu \nu} \,.
\end{equation}
In the case of a perfect fluid, for example, the physical stress-energy tensor equals $T_{\mu \nu}=(\rho+p) u^{\mu} u^{\nu}+p g^{\mu \nu}$ with energy-density $\rho$, pressure $p$ and four-velocity $u^{\mu}$. Then, the corresponding energy density and pressure in the Einstein frame are given respectively by $\rho^{\rm E} = \phi(\varphi)^{-2} \,\rho$ and $p^{\rm E} = \phi(\varphi)^{-2} \,p$.

Varying the action Eq.~\eqref{action_E}, the equations of motion read
\begin{equation}
  \Box^{\rm E} \varphi=\frac{d V(\varphi)}{d \varphi} -\frac{\beta}{M_{\rm Pl}} T^{{\rm E},\, {\rm M}}, 
\label{EOM}
\end{equation}
\begin{equation}\label{eq:einsteineq}
G_{\mu \nu}^{\rm E} = 8 \pi G \left( T^{{\rm E},\, \varphi}_{\mu \nu} + T^{{\rm E},\, {\rm M}}_{\mu \nu} \right)\,, 
\end{equation}
and 
\begin{equation}\label{eq:consstress}
\nabla^{\rm E}_\mu T_{{\rm E},\, {\rm M}}^{\mu \nu} = - \frac{\beta}{M_{\rm Pl}}  \ T^{{\rm E},\, {\rm M}} g_{\rm E}^{\mu \nu} \partial_\mu \varphi \,,
\end{equation}
where we consider a matter stress tensor $T_{{\rm E},\, {\rm M}}^{\mu \nu}$ consisting of a perfect fluid. The initial data consist of cold stars described by a polytropic equation of state $p/c^2 = K \rho^\Gamma$ with $\Gamma =2$ and $K=123 G^3 M_{\odot}^2 / c^6$. In the following, we will exclusively work in the Einstein frame and omit the indices ``$E$'' denoting the Einstein frame for the sake of clarity.

\section{Simulating neutron star mergers in $R^2$ Gravity
\label{sec:numerical}}

In this section, we go beyond a Newtonian analysis of binary mergers in theories with finite-range scalar forces. A complete treatment of the fully dynamical strong-gravity regime requires us to give up on full generality and choose a particular framework. We focus on the $R^2$ gravity model described in Sec.~\ref{sec:r2gravity}, performing fully relativistic hydrodynamic simulations of individual neutron stars and binary neutron star mergers. This model has the benefit of having only one free parameter, $a_2$, which directly controls the mass of the scalar degree of freedom in the Einstein frame via Eq.~(\ref{mR2}), and therefore the range of the scalar force sourced by unscreened matter distributions. The numerical techniques we use have been developed and tested in~\cite{Anderson:2006ay,Palenzuela:2006wp,Liebling:2002qp,Anderson:2007kz,Anderson:2008zp,Barausse:2012da,had_webpage,lorene_webpage}. We refer the reader to these references, which describe the numerical implementation of the system of Eqs.~\eqref{EOM}, \eqref{eq:einsteineq}, and \eqref{eq:consstress}. 
Briefly, and for reference, we employ a finite difference approximation of the equations of motion, based on the Method of Lines, on a regular Cartesian grid. We adopt fourth order accurate spatial discretization 
satisfying the summation by parts rule, and a third order accurate (Runge-Kutta) time integrator which
provides a robust approach to achieve stability of our numerical 
implementation~\cite{cal,Calabrese:2003yd,ander}. We employ adaptive mesh refinement (AMR) via the \textsc{had} computational infrastructure which provides distributed, Berger-Oliger style AMR \cite{had,lieb} with full sub-cycling in time, together with an improved treatment of artificial boundaries \cite{lehner}. Each grid evolves with a time step satisfying
$\Delta t_{l} = 0.25 \, \Delta x_{l}$ to guarantee satisfying the Courant-Friedrichs-Levy condition. In all cases we employ
the \textsc{Lorene} library~\cite{lorene_webpage} to define consistent initial data in GR describing isolated neutron stars
or binary neutron stars in a quasi-circular regime. In all our simulations, we monitor
the constraint violations and confirm they stay roughly at the level of that measured
at the initial time throughout our evolutions. As well, we have verified convergence in a few
representative cases in the case of single stars discussed next (since they are less computationally demanding than binaries).

With this implementation, we focus first on the study of individual stars and then binary neutron star systems. In particular,
in Sec.~\ref{sec:individualstars} we explore the properties of individual neutron stars in the Einstein frame of $R^2$ gravity as a function of $a_2$. We empirically determine the scalar charge as a function of $a_2$ from the scalar profiles sourced by neutron stars, and describe how the scalar profile can exhibit long-lived dynamical excited states. In Sec.~\ref{sec:nsmergers}, we perform a comparative study of binary mergers in GR and $R^2$ gravity sourcing both a long-range and short-range scalar force. These simulations confirm the qualitative picture of inspiral described in Sec.~\ref{Sec:Newton}, and reveal novel effects in the post-merger waveform that are of potential utility in constraining theories giving rise to short-range scalar forces. 

\subsection{Neutron stars in $R^2$ gravity}\label{sec:individualstars}

To determine the properties of individual neutron stars in $R^2$ gravity, we perform a suite of fully dynamical simulations of non-rotating neutron stars. 
Our initial data consists of a fluid that would, within GR, describe non-rotating, 
static spherically symmetric solutions (constructed with \textsc{Lorene}). Here, we supplement
this by choosing the scalar field to be initially zero and so these solutions are no 
longer static (see discussion in~\cite{Goswami:2014lxa} and Appendix~\ref{sec:timedependent}).
We choose stars with ADM mass $M \simeq 1.08 M_{\odot}$, which have compaction $C \simeq 0.07$ for the polytrope considered here (see Sec.~\ref{sec:eoms}), and vary $a_2$ in the range $a_{2}=\{\unit[21.8]{km^2}, \ldots, \unit[5000]{km^2}\}$. Recall from Eq.~\eqref{mR2} that the square of the mass of the scalar degree of freedom in the Einstein frame is inversely proportional to $a_2$, and therefore smaller values of $a_2$ correspond to larger masses. The scalar field is initialized to
 $\varphi=0$, and subsequently evolves in the presence of the star. The evolution is tracked for $t=\unit[2.46]{ms}$, corresponding to roughly $20$ crossing times of the neutron star. 

For larger values of $a_2$, the scalar profile quickly grows in the presence of the neutron star, oscillating for a few cycles, and sourcing (through the scalar monopole moment) an outgoing pulse of scalar radiation. For values $a_2 \agt 10^2 \ {\rm km^2}$, the scalar radiation is quite efficient, and the scalar and density profiles quickly settle into a nearly static configuration. For smaller values of $a_2$, the scalar profile oscillates for many cycles, sourcing very little outgoing scalar radiation. There is also an associated oscillation in the density. These excited states of the scalar profile are longer lived than the simulation time, with the amplitude and frequency of oscillation growing with mass. Apparently, the oscillating profile for small $a_2$ is not a very good antenna for scalar radiation. We provide a simple illustration of this phenomenon in Appendix~\ref{sec:timedependent}. A similar phenomenon occurs in the context of oscillons~\cite{Gleiser:1993pt}: long-lived, spatially localized, oscillating configurations of a scalar field with an appropriate non-linear potential. As shown in e.g.~\cite{Gleiser:2009ys,Hertzberg:2010yz}, the hierarchy between the characteristic size of the oscillon (or more precisely, their frequency content) and the Compton wavelength of the scalar in vacuum is responsible for their long lifetimes. This matches the trend for the neutron star solutions studied here~\footnote{There are additional effects arising from the non-linearity of the scalar potential and the non-linearities due to the coupling to gravity which could be relevant here as well.}, and suggests that excited state profiles may be sufficiently long-lived to be phenomenologically relevant for scalar forces whose Compton wavelength is of order the size of the neutron star or smaller. 

\begin{figure}[t]
\centering
 \includegraphics[width=0.49\textwidth]{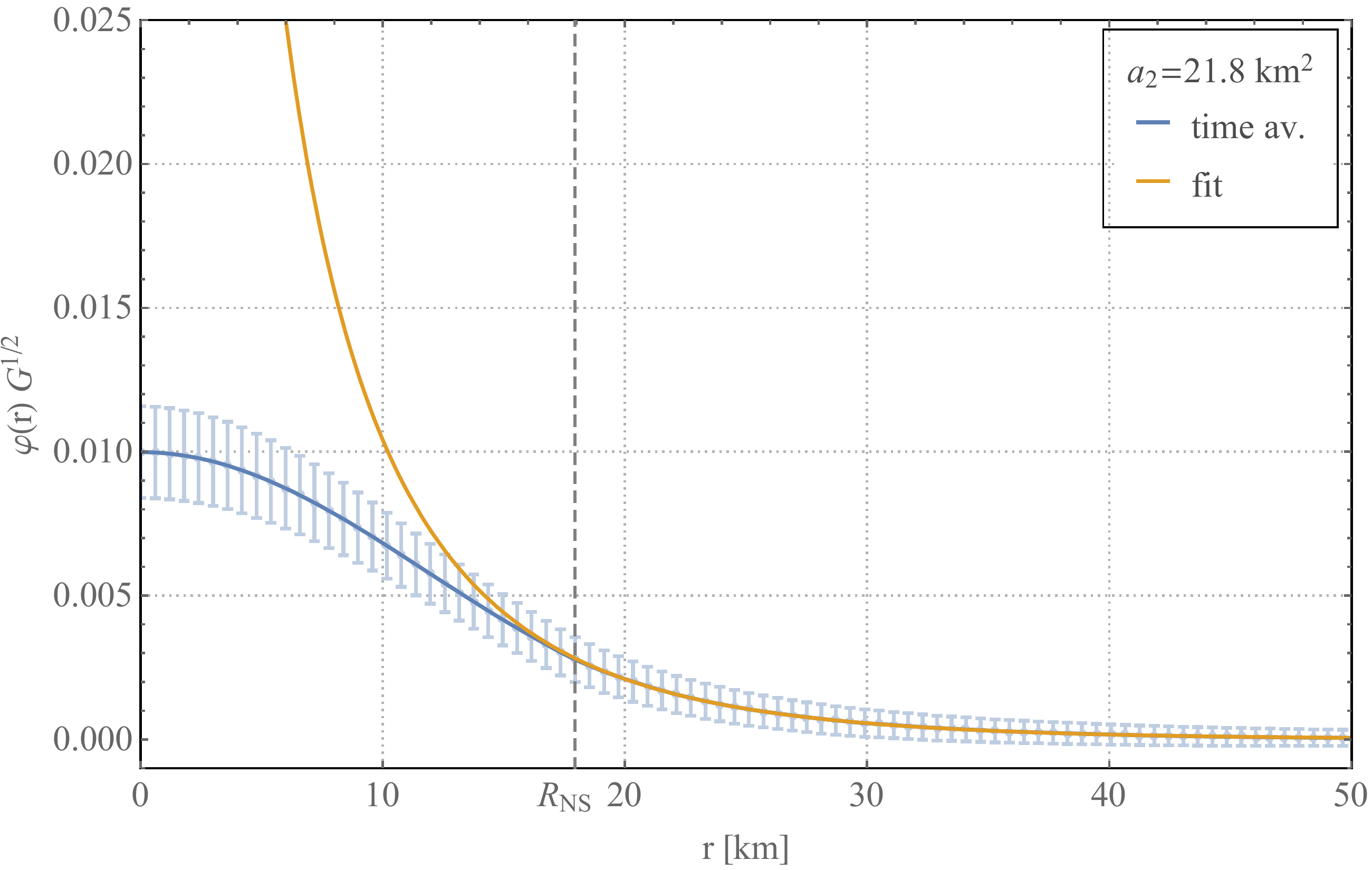}
 \includegraphics[width=0.49\textwidth]{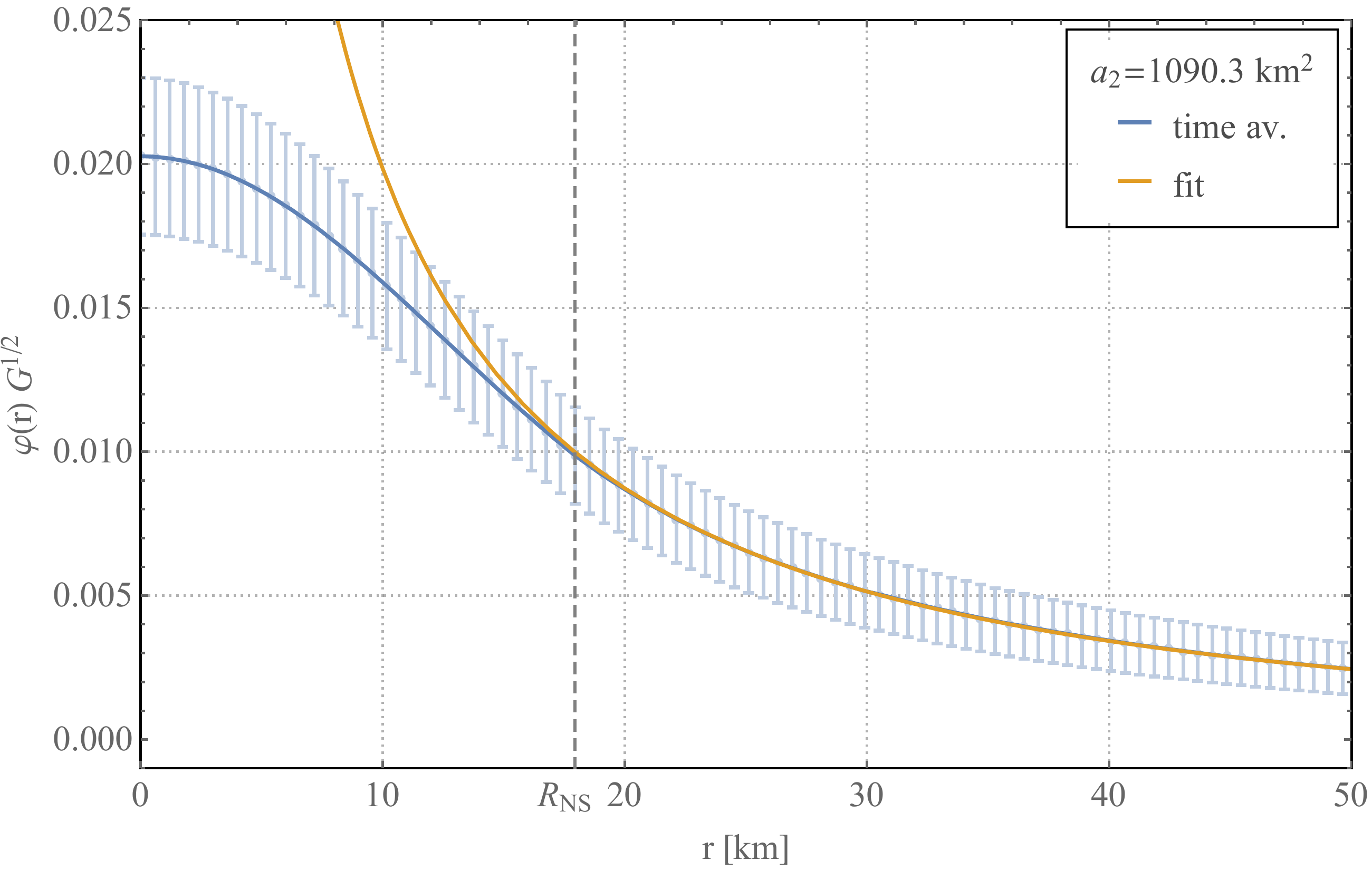}
\caption[
]
{
Time-averaged scalar field profiles $\varphi(r)$ in $R^2$ gravity for $a_{2}=\unit[21.8]{km^2}$ (left panel) and $a_{2}=\unit[1090.3]{km^2}$ (right panel). As in Fig.~\ref{rho}, the blue line displays the time average of the scalar field profiles, the light blue errors correspond to the standard deviation of the time average and the gray dashed line is the estimated neutron star radius, $R_{\rm NS}\simeq \unit[17.9]{km}$. In addition, the yellow line represents the best fit of the simulation data using the first expression in Eq.~\eqref{phisummary}. The errors of the fit are negligibly small and not visible here.}
\label{phi}
\end{figure}

Of primary interest to us here is to characterize the scalar force induced by the profile around neutron stars, by making contact with the charge and range defined in Sec.~\ref{Sec:Newton}. In order to obtain an approximation to the ground-state scalar profile, we time-average the density and scalar field profiles in each simulation. Two examples for the time-averaged density and scalar field profile, respectively, are depicted in Figs.~\ref{rho} and \ref{phi}. 

\begin{figure}[t]
\centering
 \includegraphics[width=0.49\textwidth]{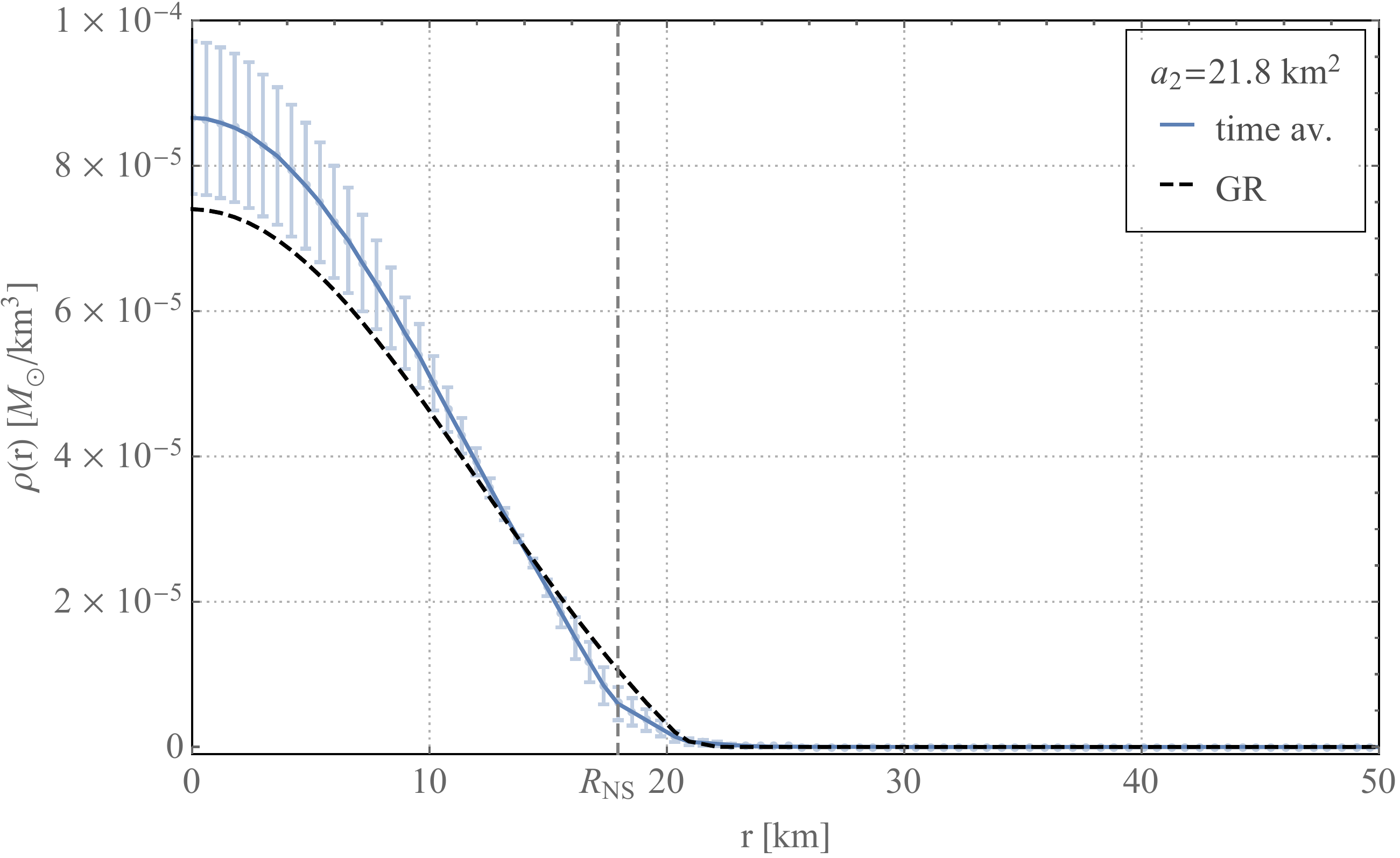}
 \includegraphics[width=0.49\textwidth]{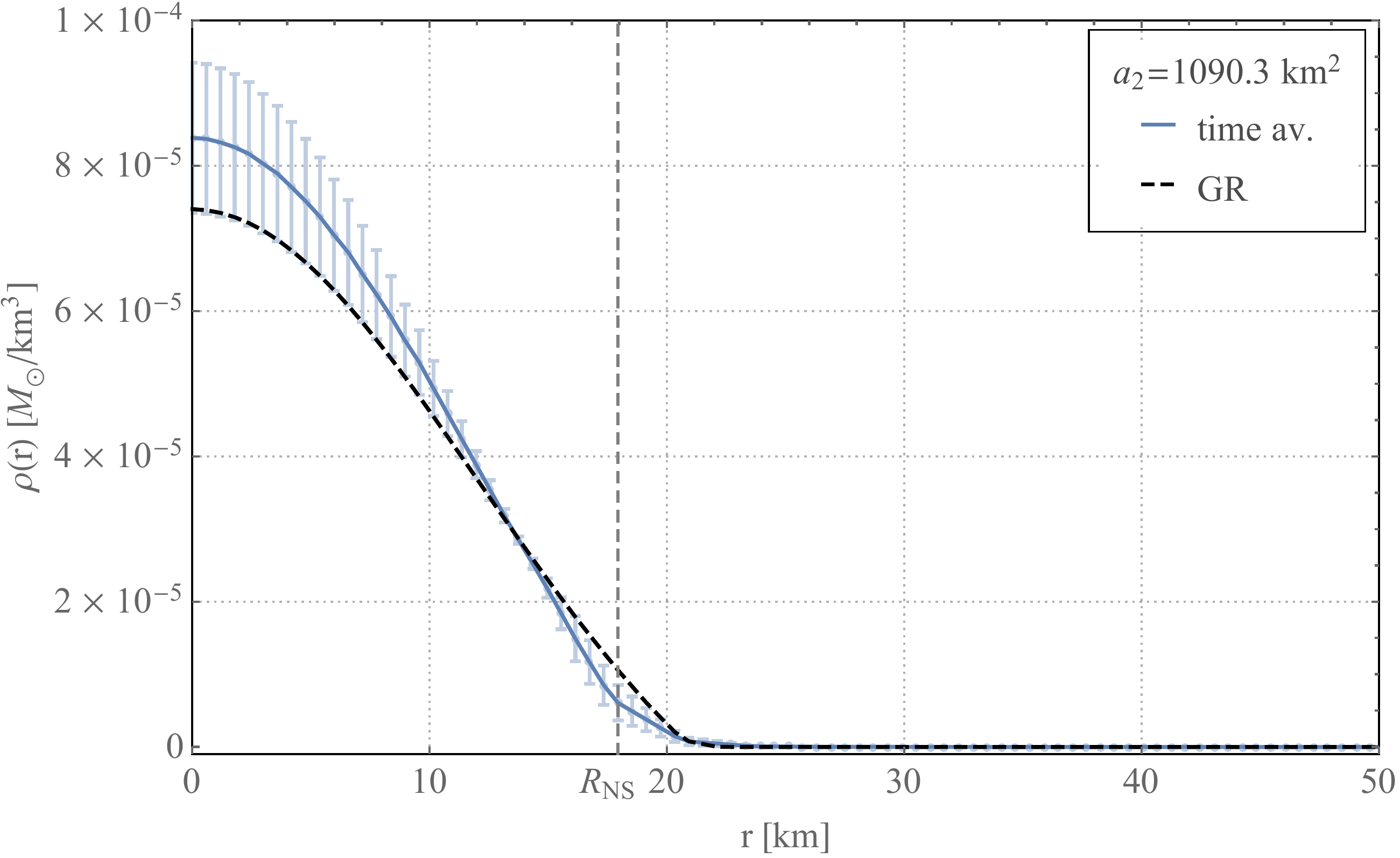}
\caption[
]
{
Time-averaged neutron star density profiles $\rho(r)$ in $R^2$ gravity for $a_{2}=\unit[21.8]{km^2}$ (left panel) and $a_{2}=\unit[1090.3]{km^2}$ (right panel). The time average of the neutron star density profile corresponds to the blue line, whereas the standard deviation of the time average is represented by the light blue errors bars. The estimated neutron star radius $R_{\rm NS}$ is indicated by the gray dashed line. In both cases, the value of the neutron star radius is approximately $R_{\rm NS}\simeq \unit[17.9]{km}$. For comparison, we also show the neutron star density profile in GR which is represented by the black dashed line.
}
\label{rho}
\end{figure}
Figure~\ref{rho} shows the time-averaged neutron star density profile $\rho(r)$ for $a_{2}=21.8$ km$^2$ (on the left) and $a_{2}=1090.3$ km$^2$ (on the right). These particular values are chosen
to span Compton wavelengths comparable to the size of the stars (left) or the computational
domain (right) and will be the ones employed in our binary simulations. While the time average of the density profile is displayed by the blue line, the error bars in light blue correspond to the standard deviation of the time average. In addition, we show the neutron star radius for each simulation (dashed gray line) estimated by the radius containing $95\%$ of the baryonic mass. Finally, we compare the neutron star density profile in $R^2$ gravity with the one in GR (black dashed line). We find that the compaction in $R^2$ gravity is in general bigger than the corresponding in GR, with compaction increasing as $a_2$ is decreased over the range of $a_2$ simulated.

The shape of the scalar field profile $\varphi(r)$ clearly varies for different values of $a_{2}$. This can be seen from Fig.~\ref{phi} where we show the time-averaged profile of the scalar field (blue line) and its standard deviation (light blue error bars) for $a_{2}=\unit[21.8] {km^2}$ in the left panel and $a_{2}=\unit[1090.3]{km^2}$ in the right panel. The strong dependence of $\varphi(r)$ on $a_{2}$ and hence on the scalar mass $m$ (see Eq.~\eqref{m}) agrees with our theoretical predictions for the scalar field. From the static solution for the scalar field in Eq.~\eqref{phisummary}, we can directly see the exponential dependence of $\varphi(r)$ on $m$. Moreover, Eq.~\eqref{phisummary} reveals the dependence of the scalar field on the parameter $\beta$, which is related to the scalar charge $\alpha$ via Eq.~\eqref{alpha}. By combining Eqs.~\eqref{alpha}, \eqref{phisummary},  we can fit the mass $m$ and the charge $\alpha$ from the scalar profile extracted from the simulations. The fit of the simulation data with this ansatz is shown by the yellow line in Fig.~\ref{phi}. Its error bars are so small that they are not visible in the plots. Clearly, the flat-space expectation for the scalar profile is a good approximation to the fully relativistic solution.

\begin{figure}[t]
\centering
 \includegraphics[width=0.49\textwidth]{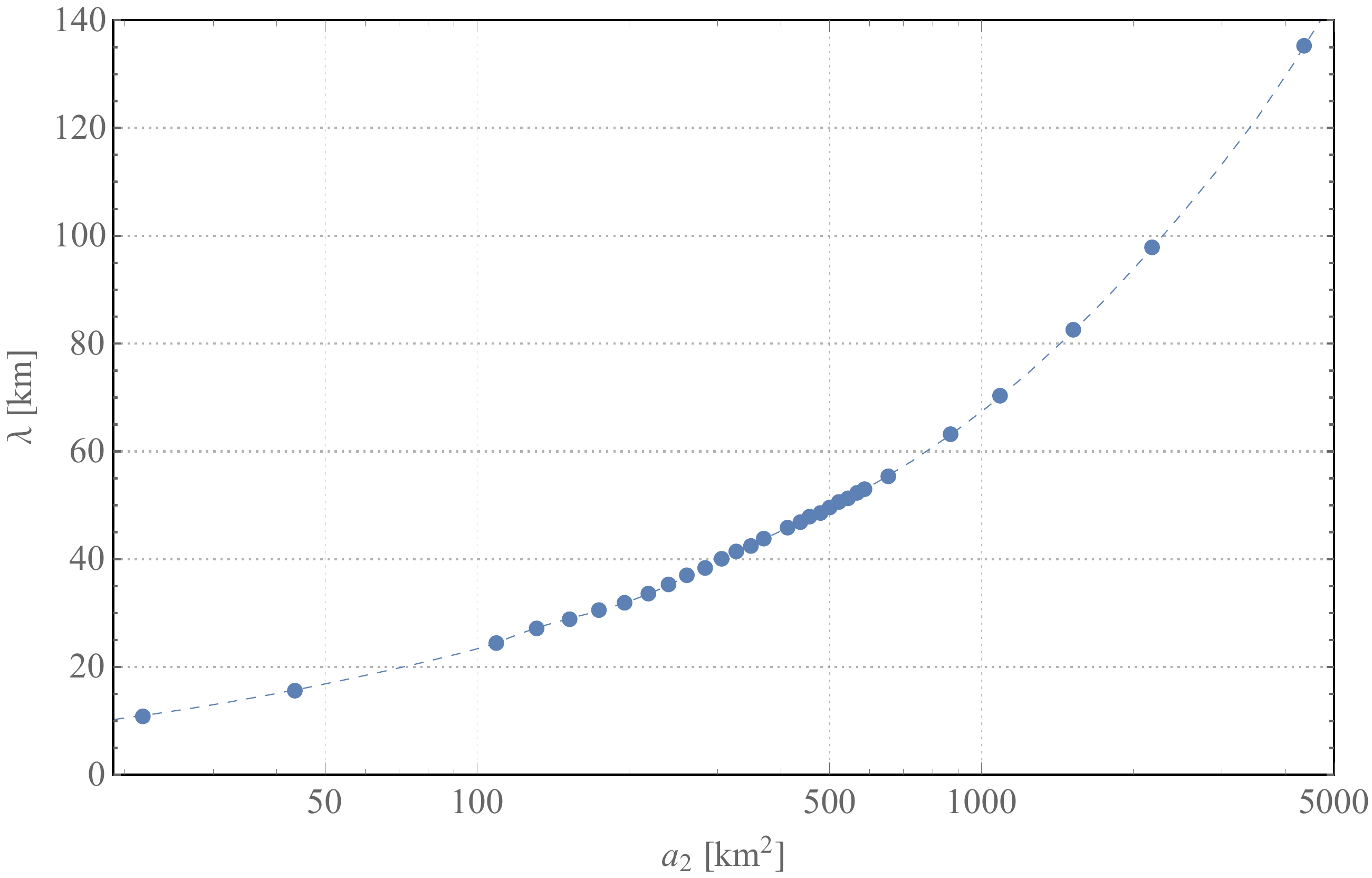}
  \includegraphics[width=0.49\textwidth]{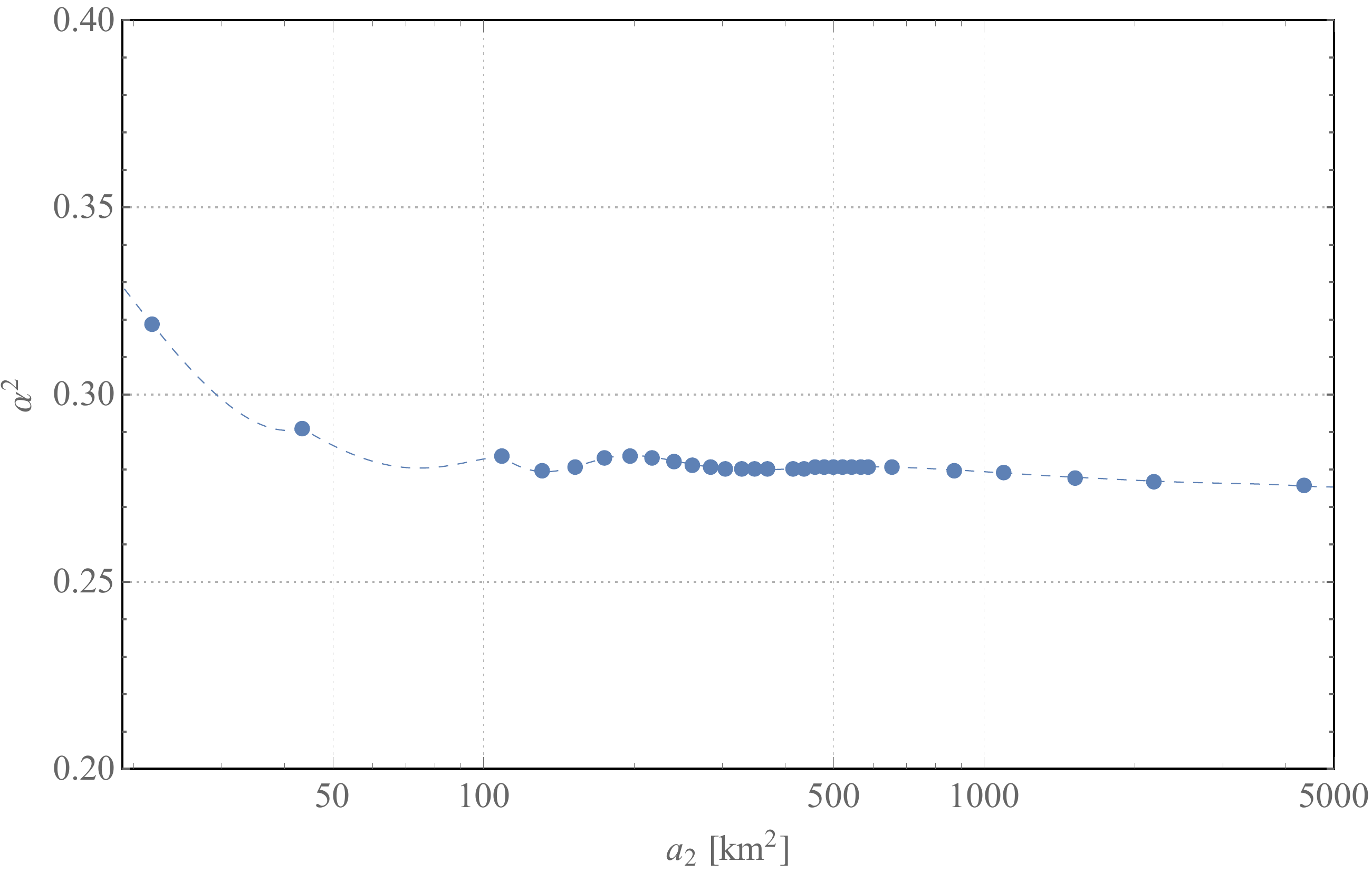}
\caption[
]
{
Compton wavelength $\lambda$ and square of the scalar charge, $\alpha^2$, as a function of $a_{2}$. The blue dots indicate the best fit parameters for $\lambda$ and $\alpha^2$ extracted from fitting the simulation data of the scalar field profiles (compare with Fig.~\ref{phi}). As in Fig.~\ref{phi}, the error bars of the best fit values are too small to be visible. The blue dashed line is an interpolation of the extracted values. Note that all profiles are averaged over a same period of time, and therefore results in different residual phases for different examples. That is why there are wiggles in the right panel.
}
\label{lambda_alpha}
\end{figure}
Results from fitting for the scalar Compton wavelength $\lambda$ and scalar charge $\alpha$ are displayed in Fig.~\ref{lambda_alpha}. We show the Compton wavelength $\lambda$ (on the left) and the square of the scalar charge $\alpha^2$ (on the right) as a function of $a_{2}$. As in Fig.~\ref{phi}, the error bars of the extracted fit parameters are negligibly small and not visible in the plots. We find good agreement between the values for the Compton wavelength of the scalar field from the simulation and the theoretical prediction (see Eq.~\eqref{m} and note that $\lambda=1/m$))~\footnote{There is some deviation at the $(1-10)\%$ level, particularly for large $a_2$. Given the quality of the fits, it is unclear what the source of this deviation is.}. For the scalar charge, we find that $\alpha^2<1/3$ for large values of $a_2$ (recall that we expect $\alpha^2 = 1/3$ for a non-relativistic star). This is to be expected because the scalar couples to the trace of the energy momentum tensor, and  the significant pressure inside the neutron star therefore hinders the ability of the scalar to couple to density. A lower coupling has the effect of lowering the scalar charge somewhat. The results from Fig.~\ref{lambda_alpha} are immediately useful for determining the constraining power from the inspiral phase as a function of $a_2$ for the binary considered here. 

\subsection{Neutron star mergers}\label{sec:nsmergers}

We now move on to describe, in fully dynamical strong gravity, coalescence using some representative 
numerical relativity simulations. 
To explore the dynamics, we focus on an equal mass binary in a quasi-circular configuration 
with masses 
$M_{1}=M_{2}=1.32 \ M_{\odot}$, described by a polytropic equation of state with $\Gamma=2$. 
This mass is not only close to inferred masses in galactic binary neutron stars system
but is also chosen to avoid prompt-collapse after merger (in GR), which yields a rotating, 
deformed massive neutron star instead lasting for at least 10ms after the stars come into contact.  
We initialize the scalar field to zero and let the
dynamics grow the field to its quasi-stationary configuration as the orbit ensues.

\begin{figure}[tbp]
\centering 
\includegraphics[width=0.4\textwidth]{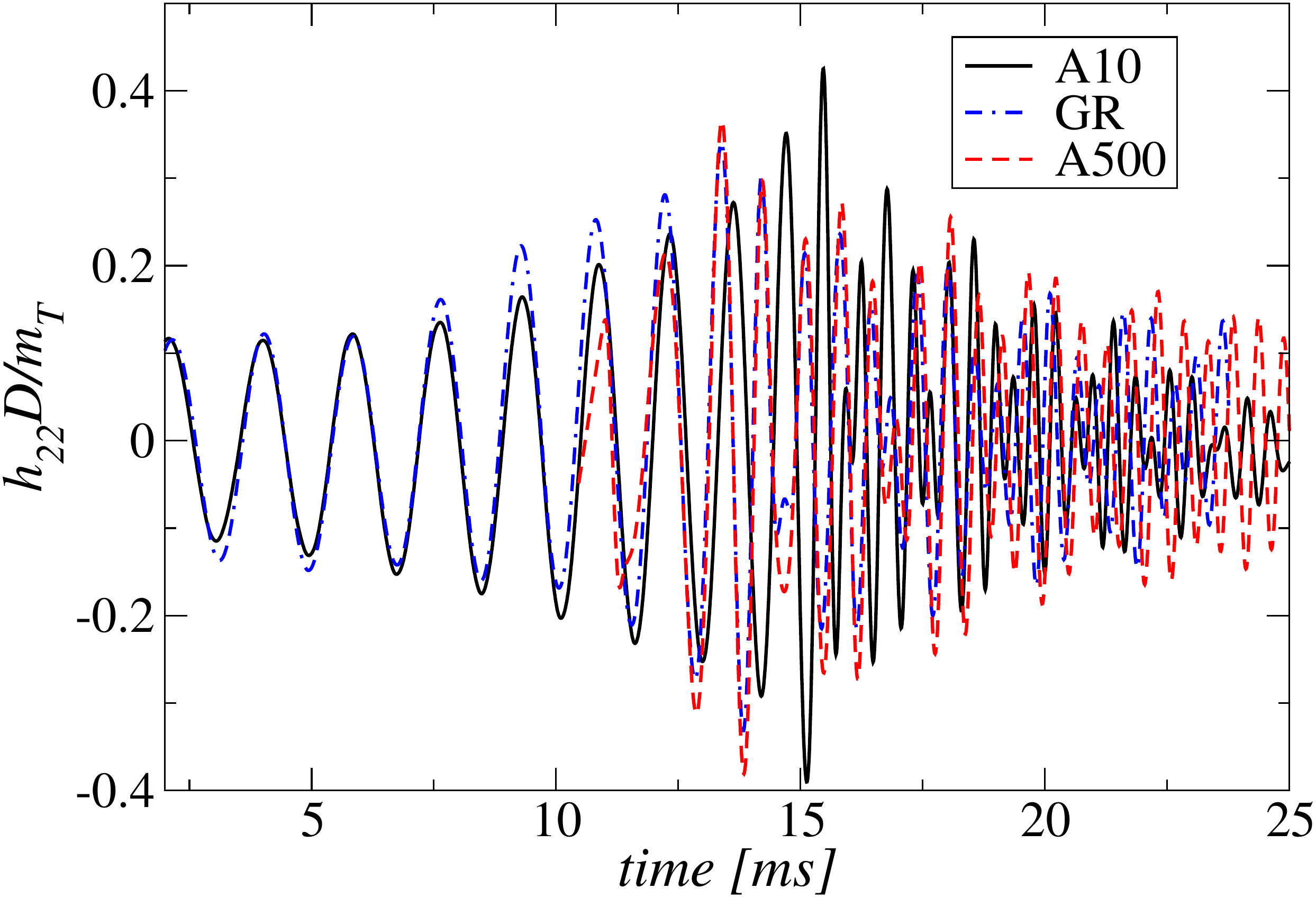}
\caption{Strain for all three cases (labeled as: A10 ($a_2=21.8 {\rm km}^2$), A500 ($a_2 = 1090.3 \ {\rm km}^2$), GR for the short-range, long-range and GR cases respectively). Notice the long-range scalar force case has been shifted
in time to ease the comparison across all cases. The stars contact each other at $t\simeq 13$ ms in the plot.\label{fig:waveforms}}
\end{figure}

\begin{figure}[tbp]
\centering 
\includegraphics[width=0.31\textwidth]{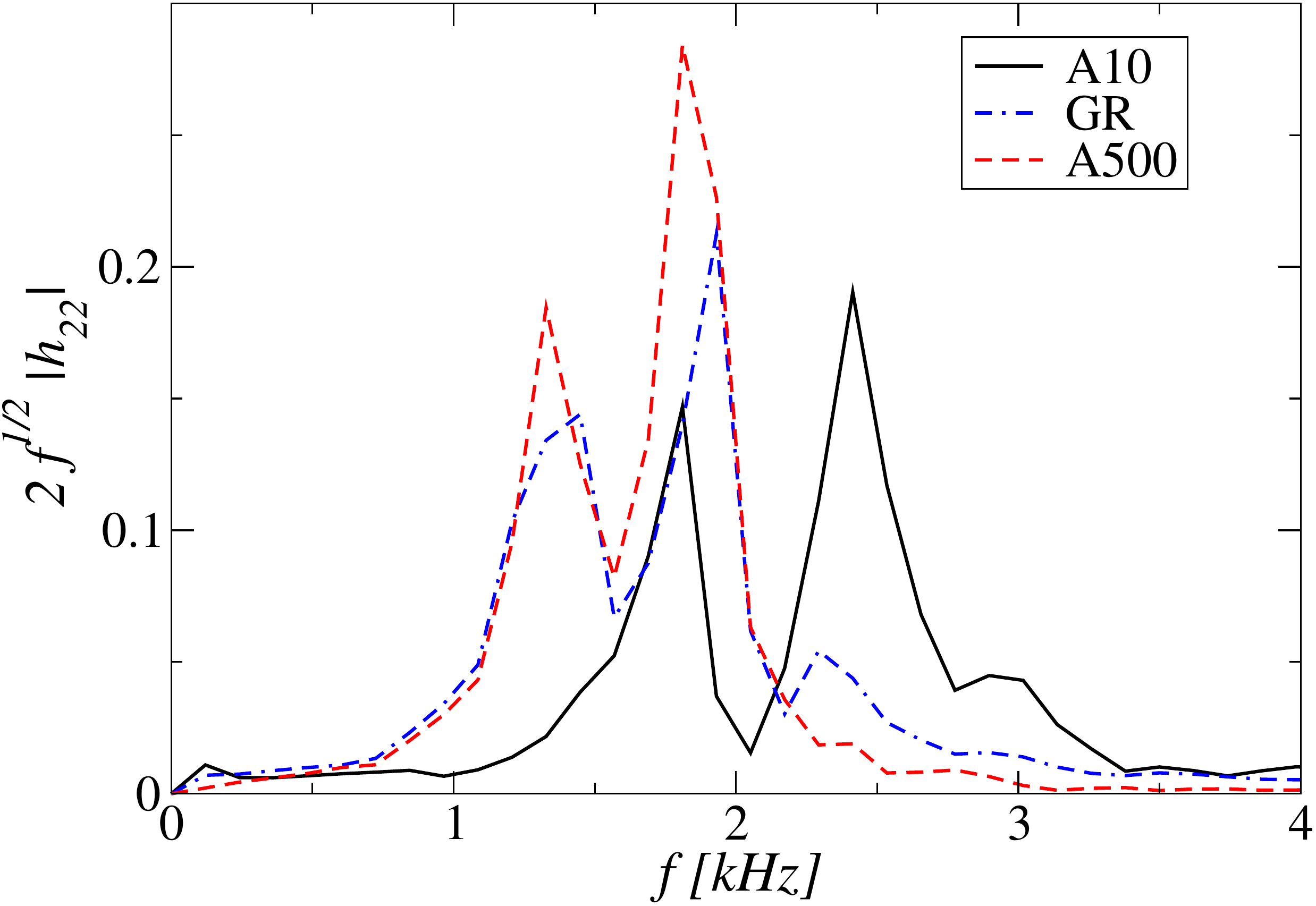}
\includegraphics[width=0.31\textwidth]{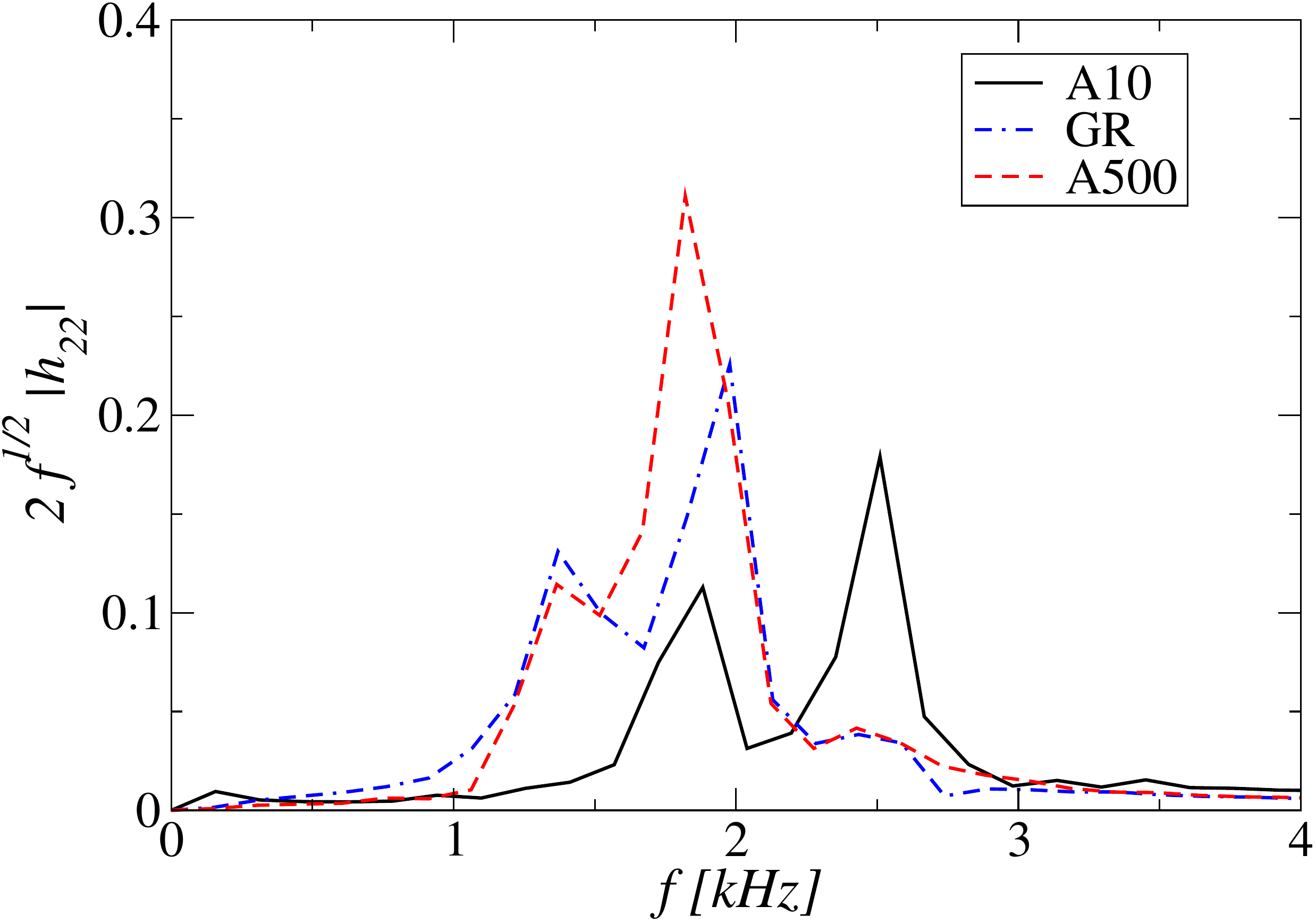}
\includegraphics[width=0.31\textwidth]{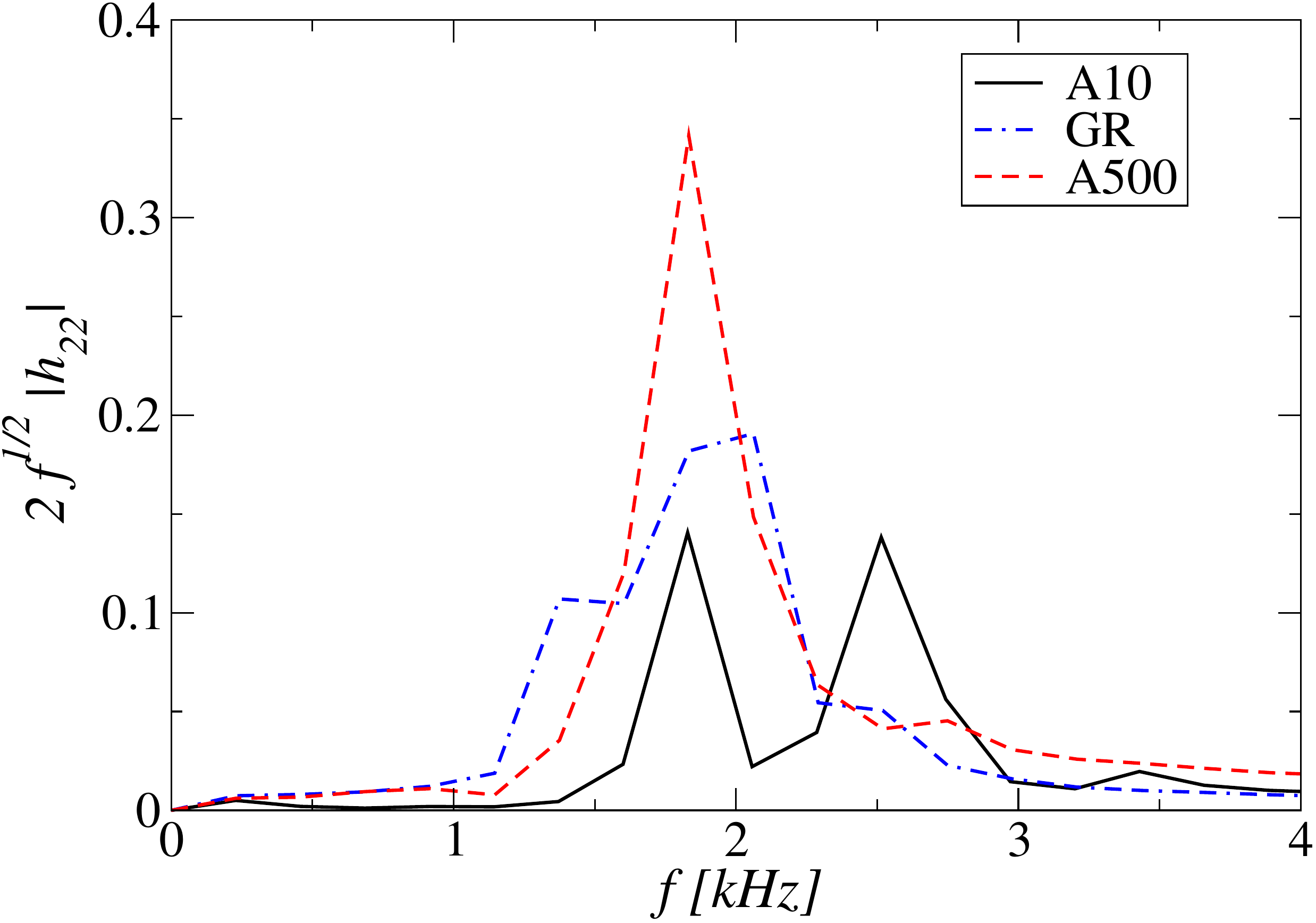}
\caption{ (Square root of the) Power spectral density (for the after-merger state) for all three cases.
The left plot is obtained by considering the time frame $[17-25]$ ms, the center panel spans $[19-25]$ ms, and the right panel spans $[21-25]$ ms.
\label{fig:waveforms2}}
\end{figure}

We first simulate the merger in GR, initializing the binary at a separation of $50$ km, which yields 
 $4$ orbital cycles before the merger, and tracking the post-merger evolution for at least $10$ ms. We extract the Newman-Penrose scalar $\Psi_4 = \ddot{h}_+ - i\ddot{h}_{\times}$ on shells in the wave zone, and integrate 
in time to obtain the strain. The result is shown as the dashed-dotted blue line in  Fig.~\ref{fig:waveforms}, which displays a characteristic chirp during the inspiral followed by a rich post-merger waveform.
In  Fig.~\ref{fig:waveforms2}, we show the post-merger
power spectral density (PSD) of the waveform. In the left panel, we choose a time window defined after the peak of strain (which takes place $\simeq 2$ ms after contact) until the end of our simulations in the time frame $[17-25]$ ms; in the center panel we choose the time frame $[19-25]$ ms to remove some of the earlier transient in the merger; in the right panel we show the time frame $[21-25]$ ms corresponding to the last part of the simulation. As seen in previous work 
(e.g.~\cite{Bauswein:2015yca,Bernuzzi:2015rla,Lehner:2016lxy,Lehner:2016wjg,Hanauske:2016gia}), there are two clear peaks. The higher frequency, larger amplitude peak, at $\simeq 1.89$ kHz is associated with the quadrupole moment of the massive neutron star resulting from the merger, and in GR it has been shown to be correlated to roughly twice the orbital frequency at the time of contact. Indeed, using the fit presented in~\cite{Lehner:2016wjg}, the
estimated frequency is $1.93$ kHz, in very good agreement with the measured value. The lower frequency, lower amplitude peak arises from the radial modes of the neutron star and/or spiral deformation induced after the merger~\cite{Bauswein:2015yca,Bernuzzi:2015rla,Takami:2014tva} though as time progresses its relevance is significantly reduced (e.g.~\cite{Foucart:2015gaa}). This can be seen comparing the three panels of Fig.~\ref{fig:waveforms2}.

We perform two simulations in $R^2$ gravity for the same neutron star system described above, choosing a value of $a_2 = 1090.3 \ {\rm km}^2$ that yields a scalar force whose range is of order the size of the simulation box and a value of $a_2 =  21.8 \ {\rm km}^2$ that yields a scalar force whose range is of order the size of the neutron stars. The scalar field is initialized to zero, but ``turns on'' in a short time
scale during the small portion of an orbit. As described above, this initial condition leads to an initial burst of scalar radiation associated with the monopole moment of each star, followed by a period during which the scalar profile around each star oscillates in time. The gravitational wave forms are extracted as described above, and plotted in Fig.~\ref{fig:waveforms} in dashed-red (long-range scalar force) and solid-black (short-range scalar force). The waveform for $a_2 = 1090.3 \ {\rm km}^2$  has been shifted in time such that the peak strain occurs at the same moment as for the GR case. In Fig.~\ref{fig:waveforms2}, we show the post-merger component of the waveforms in the frequency domain in dashed-red (long-range scalar force) and solid-black (short-range scalar force) lines.

In the case of a long-range scalar force, the merger is clearly accelerated as compared to GR. In particular
the sweep upwards in frequency and amplitude is rather rapid. For instance, the strain amplitude and frequency essentially
doubles in just a single orbit. There is also a clear modulation of the frequency.
This behavior is determined by several factors. First, the orbit is initialized with an angular velocity that is consistent with a circular orbit in the presence of the ``GR'' gravitational force alone. Therefore, the contribution to the net gravitational force provided by
the scalar force, which is activated soon after the simulation begins, induces a nearly radial impulse that accelerates the merger process.
Second, there is a faster rate of orbital decay due to a higher luminosity of gravitational radiation associated with the higher angular frequency at fixed separation which is evidenced in the more rapid/intense sweep in the chirp during the coalesence as indicated in Fig.~\ref{fig:waveforms}. These factors prevent a direct comparison with the inspiral model presented in Sec.~\ref{Sec:Newton}. However, the qualitative picture is maintained: the scalar force has a significant impact on the orbital dynamics.

In the case of a short-range scalar force, the waveform before the objects contact at $t \sim 13$ ms is quite near that of GR. This is to be expected, since the Compton wavelength of the scalar, $\lambda = \sqrt{6 a_2} \sim 11 \ {\rm km}$, is slightly smaller than the radius of the neutron stars. There are, however, some small deviations during the inspiral, which can be traced back to the initialization of the scalar field during the first few cycles. As the scalar field is grows, it overshoots its static profile and begins oscillating. This temporarily increases the range of the scalar force, and the resulting radial impulse makes the orbit slightly elliptical.  

As the stars merge, they form a massive neutron star which rotates at a varying frequency as the object compresses/decompresses from the perspective of the co-rotating frame. The difference in structure between the neutron stars in GR and the short range scalar force is already apparent in the waveform, with the waveform attaining a peak amplitude and frequency at a somewhat later time than for GR. This is because, as we found in Sec.~\ref{sec:individualstars}, the core of the neutron star in the presence of a short range force is somewhat more dense than in GR. The full post-merger behaviour can be more easily analyzed in the frequency domain by inspecting the (square root of the) power
spectral density as shown in Fig.~\ref{fig:waveforms2}. There is a clear shift towards higher frequencies in the case of the short-range scalar force as compared to GR, implying that in this case the scalar force inside the neutron star plays a non-trivial role in the post-merger dynamics. In addition, the dominant peak for the short-range scalar force drifts to higher frequency with time, which is consistent with the presence of a short range force accelerating the orbital frequency of the peaks in density inside the post-merger neutron star.

Examining the subleading peak, the amplitude in the case of GR and the long-range scalar force decreases with time, becoming nearly invisible in the right panel of Fig.~\ref{fig:waveforms2}. However, this is not the case in the binary with a short-range scalar force at least for about $12$ms after merger. Here, the reported oscillations in
the scalar field --which induce non-trivial variations (at the level of $10\%$) in the central density of the merged object when
compared to the GR case-- seemingly helps to maintain this subleading mode. This behavior thus represents a possible smoking gun signal for the presence
of such a short-range additional force. Further studies of this behavior are required to fully understand its generality.  At this point we find it important to stress how relevant
having some clear evidence for a possible short-range scalar force acting in the system is.
Namely, we have already discussed that pre-merger gravitational wave signals in this case 
will be quite similar to those obtained in the GR case. As discussed, the shift in the PSD
could hint of such an effect but could potentially be degenerate with equation of state effects. For instance,
employing the fit between peak frequency and contact frequency from~\cite{Lehner:2016lxy},
one can check that a binary systems described by stars of the same mass but a compaction
higher by $4\%$ would have a peak gravitational wave PSD in GR at the peak frequency we
observe. 

Interestingly there is no frequency shift, and only a moderate difference in the relative peak heights, between the long-range scalar force case and GR. 
While we note again that the behavior in this case is partially obscured by the
initial data issue discussed earlier, the fact that the post merger frequencies align well with the GR case indicates the physics after the merger
is likely robust. In the case of the short-range scalar force, the post-merger object achieves
a higher degree of compaction as it rotates and oscillates (in the comoving frame), which naturally implies that a higher frequency is achieved.
On the other hand, the long-range scalar force does not yield a more compact post-merger object, displaying a dynamics quite similar to that in the GR case.

Both the frequencies and amplitudes of peaks in the frequency domain carry useful information about scalar forces whose range is smaller than the size of the neutron star, providing a possible observable to extend the constraints from inspiral presented in Sec.~\ref{sec:forecast} to smaller masses. The likelihood of extracting such information in the post-merger stage, even in the case of GR alone,
has been the subject of increased scrutiny recently, 
e.g.~\cite{Bernuzzi:2015rla,Bauswein:2015yca,Radice:2016rys,Foucart:2015gaa,Lehner:2016wjg}. One can
readily extend the knowledge drawn in such studies to the question of extracting, for instance, the peak mode frequency of the post-merger
oscillation. In the case of individual events, it will be difficult to extract such information unless a ``fortunate'' event happens sufficiently
close, e.g. within $50$ Mpc, the mode lasts for sufficiently long (in the order of tens of milliseconds) and a rather aggressive (SNR $\simeq 3$) 
threshold is adopted. Otherwise, statistically, one will require a more sensitive detector (like the Cosmic Explorer or the Einstein Telescope)
and a suitable combination of several events~\cite{Yang:2017xlf}).

\section{Conclusions}
\label{Sec:conclusions}

In this paper, we have demonstrated that the observation of neutron star mergers with existing and near-future gravitational wave observatories can provide powerful constraints on finite range scalar-forces, which appear in many modifications of General Relativity. A set of high signal-to-noise detections by Advanced LIGO could tightly constrain scalar forces with a range of order the size of individual neutron stars, $\lambda \sim \mathcal{O}(5-10 \ {\rm km})$, using the inspiral phase alone. Focusing on a particular theory of modified gravity giving rise to a finite-range scalar force, the $R^2$ model of $f(R)$ gravity, a suite of numerical simulations shows that there is additional information in the post-merger waveform that could be targeted in future gravitational wave observatories to provide constraining power for forces with an even smaller range. 

The analysis we have presented here motivates a more systematic study of constraints on theories with finite-range scalar forces such as $f(R)$ gravity. Several possible future directions include:
\begin{itemize}
\item A study of models where non-linearities in the additional scalar degrees of freedom play an important role. For the particular model of $f(R)$ gravity we have chosen, the additional scalar degree of freedom in the Einstein frame never experiences large enough field excursions to sample the non-linear nature of its potential. This need not be true in other $f(R)$ models, or more generally, for theories with scalar degrees of freedom. Non-linearities in the potential can lead to interesting dynamics in the scalar sector, such as long-lived oscillating field configurations (oscillons and oscillatons), domain walls, and runaway behavior to large field excursions. These behaviors could lead to new phenomenology during the inspiral, merger, and post-merger phases of binary evolution.
\item A more systematic study of the inspiral phase including post-Newtonian corrections and scalar radiation. In Sec.~\ref{Sec:Newton}, we presented a simplified computation of the gravitational waveform during inspiral, whose accuracy should be improved by adding post-Newtonian corrections and incorporating the (small) energy loss to scalar radiation. These corrections could play an important role in breaking parameter degeneracies (e.g. as in GR, where 1PN order corrections break the degeneracy between the individual masses in a binary merger), and may provide sensitivity to properties of the scalar profile around the neutron stars beyond the Compton wavelength and charge described above. 
\item Detailed forecasts in the context of the design of future gravitational wave observatories. In Sec.~\ref{sec:forecast}, we provided a forecast for Advanced LIGO based on the inspiral waveform alone. This analysis did not consider the full parameter space (e.g. spin, tidal effects, etc.), and worked at Newtonian order in the inspiral waveform. In addition, we did not consider the information contained in the merger and post-merger phases, which could be provided by a larger suite of numerical simulations (although this will be computationally expensive). A more detailed forecast would go beyond these simplifications, and also provide a systematic study of which experimental configurations could yield the optimal constraints.
\item Exploration of degeneracies with the neutron star equation of state. In Sec.~\ref{sec:nsmergers}, we have seen that the dynamics of the scalar degree of freedom in $R^2$ gravity can lead to important differences in the post-merger waveform. However, the post-merger waveform is also sensitive to variations of the equation of state of the neutron star. It is clear that changing the equation of state can lead to, at least in part, similar phenomena (e.g. changing the peak structure of the power spectrum as in Fig.~\ref{fig:waveforms2}). A more systematic investigation should be performed to understand what further information can help break
possible degeneracies.
\item Connections with models of dark energy and screening mechanisms. Although the $R^2$ gravity model chosen here does not provide a model for cosmic acceleration, other choices of $f(R)$ (and other modifications of GR) do. In order for these models to be phenomenologically viable, the scalar force must be screened via the Chameleon mechanism (or other mechanisms, such as Vainstein screening or the Symmetron mechanism). Although this implies that there will be no signature from the scalar force during the inspiral phase (see e.g. Eq.~\eqref{eq:chameleonscreening}), there may be important differences in the post-merger waveform. This occurred in the $R^2$ model studied in Sec.~\ref{sec:nsmergers} with $a_2=21.8 \ {\rm km^2}$, where there were no effects during inspiral, but interesting differences from GR in the post-merger waveform. Such a connection would provide a new window on dark energy physics, possibly accessible with the next generation of gravitational wave observatories.
\end{itemize}

This paper represents only a small step in a systematic investigation of how to maximize the science return from future gravitational wave observatories. We hope that our results serve as further motivation for the community to undertake detailed studies of how broad this science return might be in terms of testing modifications and extensions of general relativity. Although in many cases the derived constraints on $f(R)$ gravity will not be competitive with existing constraints from terrestrial experiments (which rule out fifth forces on scales of order $\sim 1 \mu{\rm m}$ and larger~\cite{Adelberger:2009zz}), binary neutron star mergers can be used to place independent constraints in a completely different environment.

\acknowledgments
We thank Mustafa Amin and Eugene Lim for important conversations.
MCJ and LL are supported by the National Science and Engineering Research
Council through a Discovery grant; MS is partially supported by the STFC grant ST/L000326/1; SLL is supported by NSF under grant PHY-1607291 (LIU); DN is supported by NSF under grant 1607356 (BYU);
CP is supported by the Spanish Ministry of Economy and Competitiveness grant
AYA2016-80289-P (AEI/FEDER, UE).
LL also thanks CIFAR for support. This research was supported in part by Perimeter
Institute for Theoretical Physics. Research at Perimeter Institute is supported
by the Government of Canada through the Department of Innovation, Science and
Economic Development Canada and by the Province of Ontario through the Ministry
of Research, Innovation and Science.

\appendix

\section{The long-lived oscillating configuration of the scalar field}\label{sec:timedependent}

As mentioned in \ref{sec:individualstars}, we start the single neutron star simulation with $\varphi = 0$, and expect that $\varphi$ will approach a static profile by radiating energy away until the configuration settles into a stable, static state. However, we find that there still are some oscillating modes left, and are longer lived than the simulation time. In this appendix, we try to understand why such modes persist.

Again, we assume a spherically symmetric flat spacetime and keep only the quadratic term of the scalar potential for simplicity. In this case, the time dependent equation of motion of $\varphi$ is 
\begin{equation}\label{fullEoMvarphi}
-\frac{d^2\varphi}{dt^2}+\frac{d^2\varphi}{dr^2} + \frac{2}{r}\frac{d\varphi}{dr} = m^2 \varphi + \frac{\beta}{M_{\rm Pl}} T^{\mu}_{\mu} \,.
\end{equation}
 We assume the scalar field will approach a static profile $\varphi_0(r)$ which satisfies
\begin{equation}
\frac{d^2\varphi_0}{dr^2} + \frac{2}{r}\frac{d\varphi_0}{dr} = m^2 \varphi_0 + \frac{\beta}{M_{\rm Pl}} {T_0}^{\mu}_{\mu} \,.
\end{equation}
and rewrite $\varphi(t,r)$ as
\begin{equation}
  \varphi(t,r) = \left[1+f(t)\right]\varphi_{0}(r)  + \delta \varphi(t,r) \,.
  \label{psirt}
\end{equation}
Here $f(t)$ represents the oscillations on top of $\varphi_0$. To demonstrate, we could focus on oscillations with a single frequency by choosing $ f(t)\equiv f_{0} \sin(\omega_{0} t)$. Inserting the ansatz \eqref{psirt} in Eq. \eqref{fullEoMvarphi} and subtracting the static part of the equation, we obtain the equation for $\delta \varphi(r,t)$,
\begin{equation}
-\frac{d^2 \delta \varphi}{dt^2}+\frac{d^2 \delta \varphi}{dr^2} + \frac{2}{r}\frac{d \delta \varphi}{dr} - m^2 \delta \varphi =f(t) J(r),
\label{fullEoMvarphi2}
\end{equation}
where
\begin{equation}
  J(r) \equiv  \frac{\beta}{M_{\rm Pl}} T^{\mu}_{\mu} - \omega_{0}^2 \varphi_0(r) \,.
\end{equation}
From Eq.~(\ref{fullEoMvarphi2}), we find that $\delta \varphi$ is sourced by $f(t)J(r)$, and therefore could drain energy from the oscillations. Using the retarded Green function
\begin{equation}
G_R\left(\omega | r-r'\right) = \frac{e^{\rm{sign}(\omega)}e^{i\sqrt{\omega^2-m^2}\, \left(r-r'\right)}}{\left|r-r'\right|}
\end{equation}
for $|\omega| \ge m$, we solve $\delta \varphi$ in the frequency domain
\begin{equation}
\delta \varphi(\omega | r) = - \frac{e^{\rm{sign}(\omega)}e^{i\sqrt{\omega^2-m^2}\,r}}{r} f(\omega) J_0,
\end{equation}
where we defined $J_0 = \int dr' r'^2 J(r')$. Then we can find
\begin{equation}
\delta\varphi(t, r) = - \frac{f_0 J_0}{r} \sin \left(\omega_0t-\sqrt{\omega_0^2-m^2}\,r\right),
\end{equation}
which carries an energy flux (averaged over one period) of
\begin{equation}
  \left< \partial_{t} \delta \varphi \, \partial_{r} \delta \varphi\right> = -\frac{2 f_{0}^2 J_{0}^2}{r^2} \omega_{0} \sqrt{\omega_{0}^2-m^2}\,.
\end{equation}
From this we see that the oscillation of the scalar field will decay slowly if $\omega_{0} \sim m$. It is indeed this behavior that we observe in our simulations (see Sec.~\ref{sec:numerical}).

\phantomsection   
\addcontentsline{toc}{section}{References}
\bibliography{References_ModifiedGravity}
\end{document}